\documentclass[twocolumn,secnumarabic,amssymb, nobibnotes, aps, prd]{revtex4-2}
\UseRawInputEncoding

\def \sla{\slash {\!\!\!}}

\usepackage{graphicx}
\usepackage{dcolumn}
\usepackage{bm}

\begin{document}


\title{Space-time approach to spontaneous symmetry breaking in the Abelian-gauge interaction}


\author{Shun-ichiro Koh}
\email[]{koh@kochi-u.ac.jp}
\affiliation{Kochi University,\\2-5-1, Akebono-cho, Kochi, Japan}


\date{\today}

\begin{abstract}
  Spontaneous symmetry breaking is studied by regarding it as a phenomenon in the eternal intermediate state due to  sequential perturbations.   The concept of the  relativistic  many-body state is applied to this intermediate state ocurring  in the collision of massless Dirac fermions. 
Time  in the  relativistic  many-body state  should evolve while maintaining  the  direction of time in each particle,   even if the particle  are  viewed  from any inertial frames.  This kinematical  requirement  leads to  spontaneous symmetry breaking in the vacuum of these states, which  gives a different meaning to the results of the Higgs model.
In this  vacuum,  massless fermion-antifermion pairs and coherent collection of gauge bosons condense, which determine each other's mass.   When a local excitation  of  the condensed gauge bosons   propagates in space, a Higgs-like boson  appears.   The effective coupling of this  Higgs-like boson to gauge bosons is calculated as a one-loop process. With  this coupling,  the total cross section of the pair annihilation of fermion and antifermion to gauge boson pair  is calculated.  Renormalizability of this model is discussed using the inductive method.  Since the Higgs Lagrangian is not assumed, the divergence we must renormalize is only the logarithmic divergence, not the quadratic  one.

\end{abstract}


\maketitle


\newpage 

\section{ Introduction}
\begin{figure}
	 	\centering
 \includegraphics [scale=0.35]{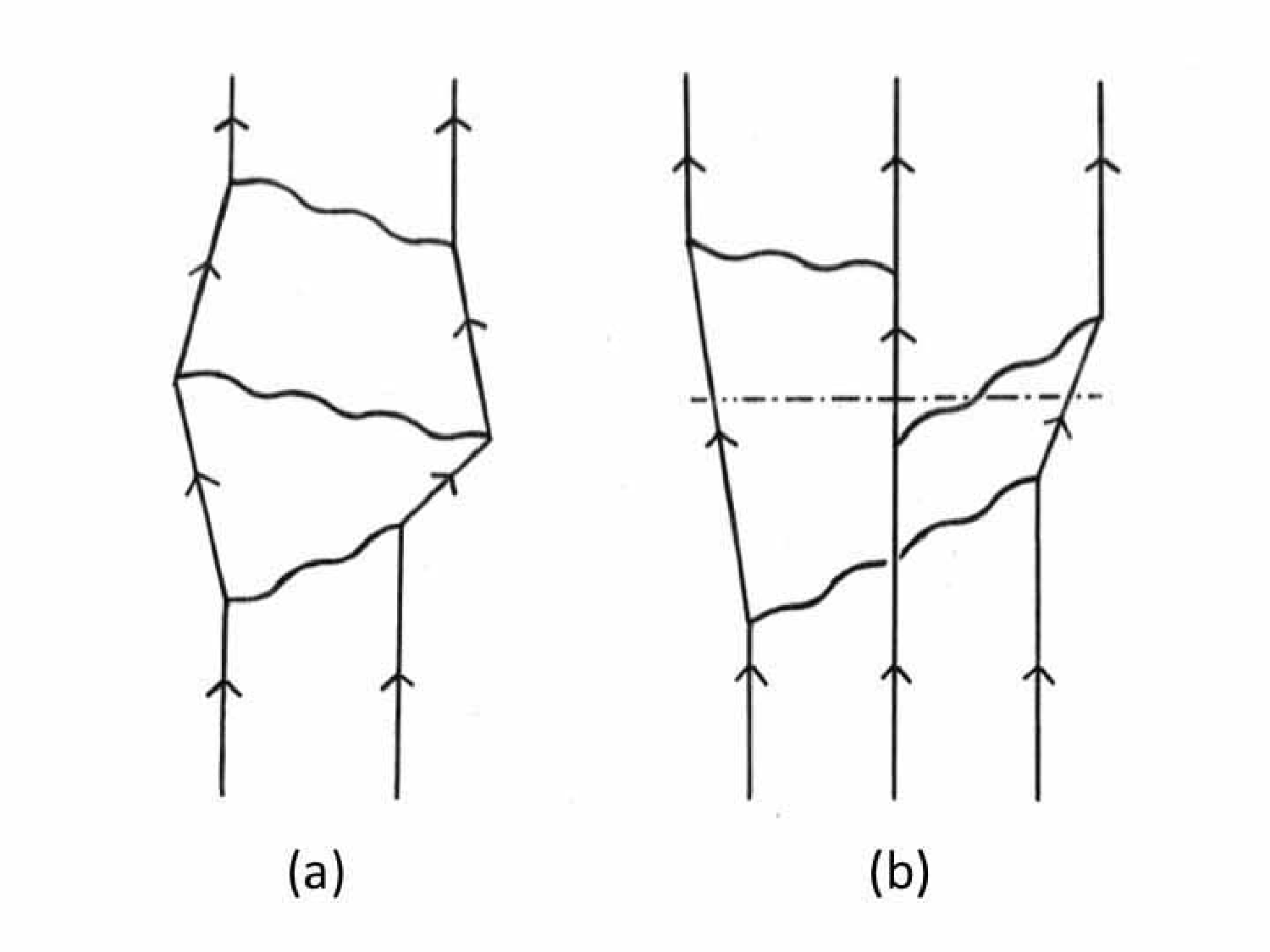}
\caption{Examples of (a) two-body  intermediate states (b)  three-body  intermediate states. }
\label{fig.01}
\end{figure}

Spontaneous symmetry breaking is examined  using a vector Abelian-gauge field $B_{\mu}$ coupled to massless Dirac  fermi field $\varphi$
\begin{equation}
		L_0(x)= -\frac{1}{4}F^{\mu\nu}F_{\mu\nu}+\bar{\varphi}(i\partial_{\mu}+gB_{\mu})\gamma^{\mu}\varphi   .
		 \label{eq:1}
\end{equation}
In the usual form of quantum relativistic field theory, the world is modeled as the physical  process that connects  initial and final asymptotic states of the particle collision. The Fock state created on the free vacuum is used for this asymptotic state. This Fock state exists only when all interactions are switched off, and this is a good approximation  in the case of short-range interaction. However, spontaneous symmetry breaking occurs in situations where the gauge field that causes long-range forces is still an extended object. Therefore, the interaction is not switched off even in the initial and final asymptotic states. This is a reason why the Fock state is unsuitable for deriving spontaneous symmetry breaking. 
In the case of long-range forces, there is always an interaction between the particles, and in this case the more natural point of view is as follows. {\em The real world is modeled as an eternal intermediate state due to  sequential perturbations} \cite{coh}. We are in the middle of such intermediate states, where the broken-symmetry vacuum is expected  to appear. Consider an intermediate state as shown in  Figure.\ref{fig.01} caused by the collision of massless Dirac fermion.  In order to obtain a broken-symmetry vacuum, we must think of a suitable representation of states.

In  quantum field theory with  infinite degrees of freedom, there are countless  inequivalent  representations with the same commutation relations  in a free vacuum,  but belonging  to different Hilbert spaces \cite{fri} \cite{van} \cite{miy} \cite{haa}.  The choice  of representation  must follow  natural phenomena, but the  form of  representation  is decided by the human side.

For the suitable representation of the broken-symmetry vacuum, let us  apply the concept of {\em  relativistic many-body state} to  this intermediate state \cite{nuc}.  There is an assumption about  the form taken by the many-body states: {\em Time  in the many-body state  should evolve  while maintaining the  direction of time in each particle,   even if particles  are viewed  from any observer } When it is applied to the intermediate states,  {\em time  in the relativistic many-body state  should evolve  while maintaining  the direction of time in each particle,   even if particles in the   intermediate states  are viewed  from any inertial frame. } 
As long  as it is a one-particle state,  this is merely a matter  of interpretation.  However,  in the many-body state, it becomes a serious problem.  If this is not followed,  there will be an object  coming from the future to the present  in the many-body state, and its relation to other objects coming from the past  cannot be explained. In this case, the  many-body state will be  chaotic. 
  The relationship  between different objects  is understood on the premise  that they follow  a  common direction of time \cite{law}. 
 In the intermediate  state, although particles  travel  not only time-like paths but also  space-like ones, the temporal order of events should not be reversed.   This criterion  should give an  inequivalent representations with a suitable form. However, in the relativistic many-body state, there is a situation that threatens this.

 \begin{figure}
	 	\begin{center}
\includegraphics [scale=0.30]{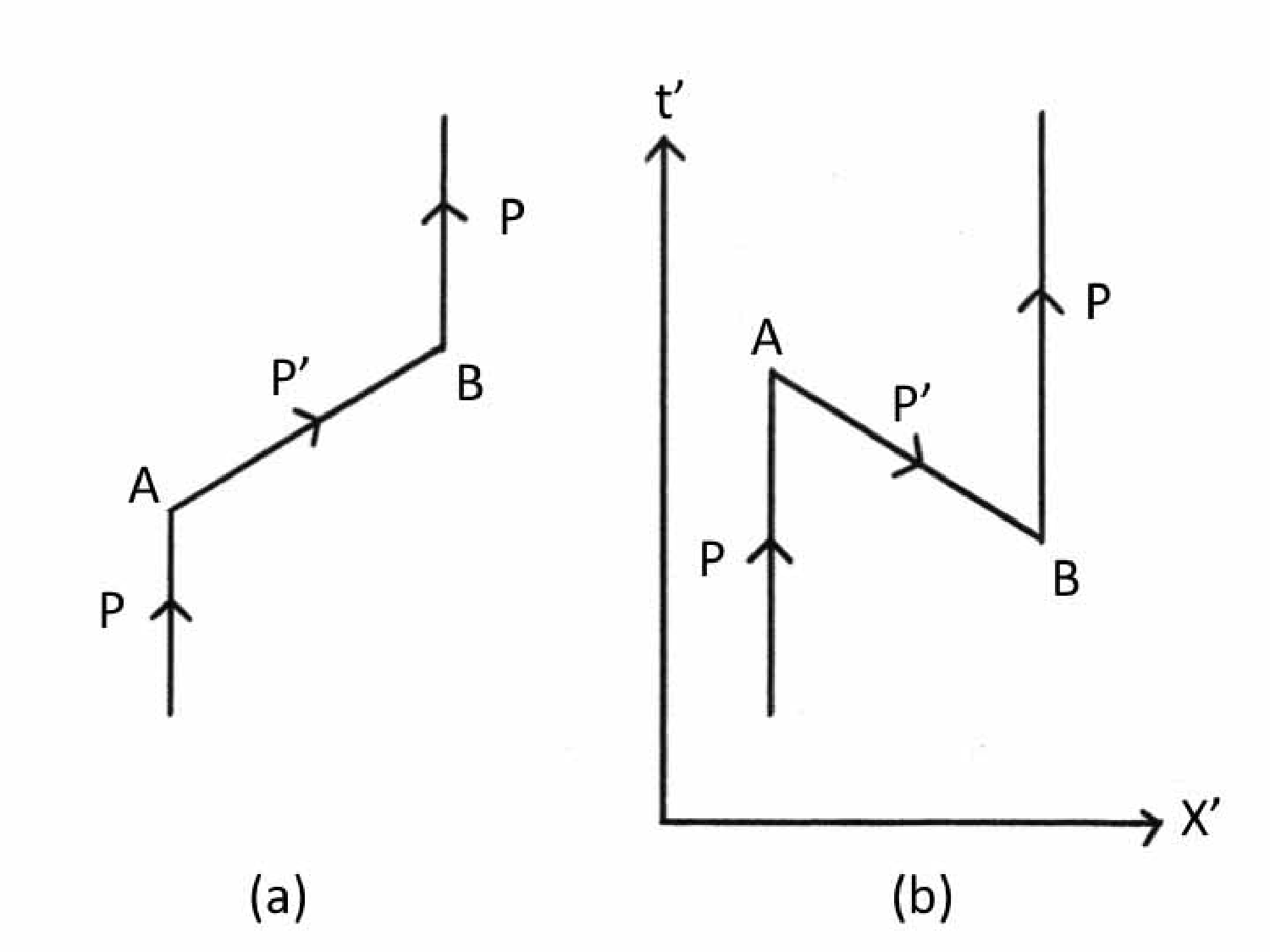}
\caption{ (a) The motion of a massless Dirac fermion  perturbed by $U_1$ and $U_2$ at $A$ and $B$  ($t_2>t_1$).  (b)  When viewed from  a fast-moving  inertial frame $(x',t')$,  the order of two events separated by spacelike interval is reversed ($t'_2<t'_1$). }
\label{fig.1}
    	\end{center}
\end{figure}

Consider the second-order perturbation process of  a moving massless Dirac fermion under disturbances in Figure.\ref{fig.1}(a). There are   two disturbances, $U_1$  in $A$ at a time $t_1$,  and $U_2$  in $B$ at a later time $t_2$, in which the second disturbance $U_2$  restore the fermion to its original state with a momentum $\mbox{\boldmath $p$}$.  Such an amplitude is calculated by  summing  all  possible, timelike or spacelike, intermediate  states between $A$ and $B$ over their momenta $\mbox{\boldmath $p$}'$ viewed from different inertial frame (see Appendix.A).  The  states on the path $AB$  look different  when viewd from other inertial systems.
 In the inertial system of a  coordinate $(\mbox{\boldmath $x$},t)$,  a fermion  with a negative electric charge and momentum $\mbox{\boldmath $p$}$ leave $A$ at $x_1$ and $t_1$,  and reach $B$  at $x_2$ and $t_2(>t_1)$.  When this motion is viewed  from another inertial system  moving in the $x$-direction  at a relative velocity $v$ to the original one, it follows a Lorentz transformation to a new coordinates $(\mbox{\boldmath $x'$},t')$.  The   time difference $t_2-t_1$ between   $A$ and  $B$ is Lorentz transformed to
\begin{equation}
		t_2'-t_1'=\frac{1}{\sqrt{1-(v/c)^2}}\left[ t_2-t_1-\frac{v}{c^2}(x_2-x_1) \right]   .
				                	 \label{eq:7}
\end{equation}
A prominent  feature of the Lorentz transformation is that  it does not leave  the temporal order of events on  the spacelike path invariant.  When the fermion  has a small velocity,   the observer has more options for having a large relative velocity $v$ to the fermion.  A  sufficiently large spacelike interval between two events,  such as $c(t_2-t_1) <(v/c)(x_2-x_1) $ in  Eq.(\ref{eq:7}),   reverses  the temporal  order of two events,  $t_2'<t_1'$ as shown   in  Figure.\ref{fig.1}(b).

The  natural interpretation  without the reversal of temporal order is that a  positively-charged antifermion runs  in the opposite direction.  
 (Historically, Stueckelberg first stressed  this  \cite{stu},  and later  Feynman independently made an intuitive  explanation for the raison d'etre of antiparticle along this line \cite{fey}.)  The intermediate many-body state should be described using antifermion  so that  the time direction is not reversed by any  observer.  Therein lies the key for  understanding the  vacuum with  broken-chiral-symmetry.

  In this paper,   we derive spontaneous symmetry breaking from a   kinematical requirement.
In Section 2, using the massless fermion  system coupled to  the Abelian-gauge field,   the concept of the relativistic many-body state is applied   to  the  intermediate states in the collision of this system, and it is shown that  the  kinematical requirement on the time direction in  this many-body state leads to the physical vacuum with broken-chiral-symmetry.
 Section 3 describes some dynamical consequences of this   kinematical requirement.  In Section 3-1, and  3-2,  the  mass  of the  fermion  is generated from gauge bosons in  the condensed field energy.  In Section 3-3, the mass generation of the  gauge boson  in the physical vacuum is explained.   In Section 3-4,  the Higgs-like boson is derived  as a local excitation propagating in the physical vacuum, the mass of which is calculated as an excitation energy. Section 4 gives an interim summary. In Section 5,  one-loop perturbation calculation is performed  for the effective  coupling of  the Higgs-like  mode to the gauge field,  and  for the self interaction of this mode.  In Section 6,   the total cross section is calculated for  the pair-annihilation of massive  fermion and antifermon   to massive gauge bosons.  In Section 7,  renormalizability of this model is examined.   In Section 8,   some generalization to the electroweak interaction is  briefly discussed. 
 
\section{Relativistic  many-body states of  massless Dirac fermions}
\begin{figure}
	 	\centering
 \includegraphics [scale=0.35]{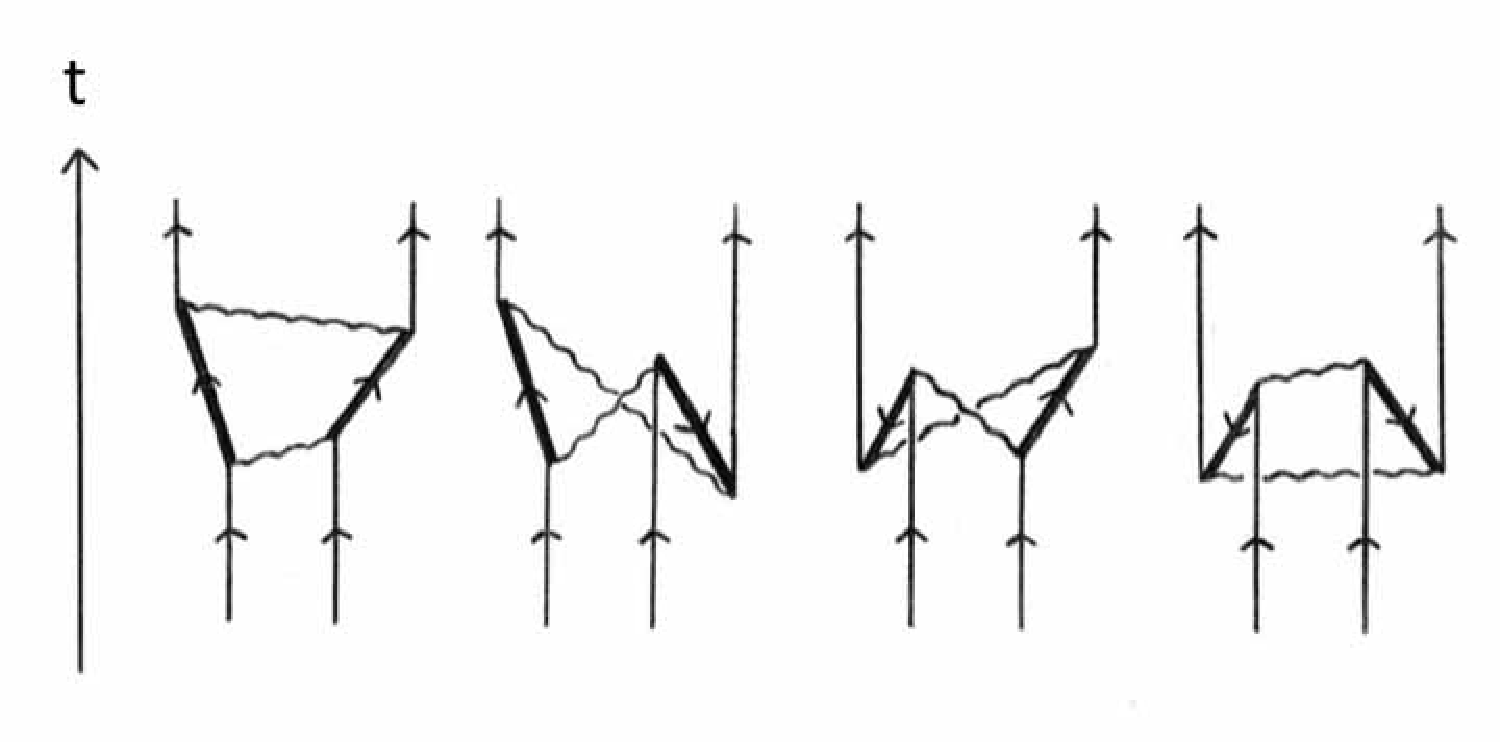}
\caption{The relativistic  two-body states appearing in the  intermediate state of  massless fermions  (thick lines with arrows).  
Four types of  combination appear on  the temporal order of two fermions, if antifermion is not introduced.  }
\label{fig.23}
\end{figure}

As an example of the relativistic  many-body state, consider  the intermediate state in  the  direct  scattering of two massless  Dirac fermions, and regard it as a two-body state.  Figure.\ref{fig.23} shows its fourth-order process.   The  particle  in the intermediate state    (thick lines with arrows) looks  a normal or time-reversed one, depending on what inertial frame it is viewed from.   Hence four different combinations of the temporal order appear  in the two-body state \cite{opp}
(The exchange scattering is also  possible,  in which the gauge-boson lines are crossed at the left and right ends of  Figure.\ref{fig.23}, and not crossed elsewhere.) 
  The non-relativistic two-body state contains only the left  end of Figure.\ref{fig.23},  whereas the  relativistic two-body state is a superposition of all possible combinations.
To regard the relativistic  intermediate state as the many-body state,  no matter what form an individual particle takes,  time should  evolve  in the same direction  there no matter what inertial frames they are viewed from.  Hence,  these time-reversed motions should  be interpreted  as the motion of antifermion.  All processes in Figure.\ref{fig.23},  the propagation of massless  fermion, the  pair creation and  annihilation of massless  fermion and antifermion,  occur in the common direction of time.
 Each vertex in Figure.\ref{fig.23} has two possibilities, a vertex of massless fermion or that of antifermion. 
 To obtain the representation which is always valid when viewed from any reference frame,  combine two pictures  in  Figure.\ref{fig.1}. 
Momentum, electric charge and spin,  which prescribe all  properties of massless fermion,  are not positive definite, and therefore  the operation of  $a^s(\mbox{\boldmath $p$})$ on the state can be equated  with that of $b^{s\dagger }(-\mbox{\boldmath $p$})$.
We define  a new annihilation operator  at $A$  in  Figure.\ref{fig.1}  by considering  a superposition of $a^s(\mbox{\boldmath $p$})$ and $b^{s\dagger }(-\mbox{\boldmath $p$})$
\begin{equation}
		A^s(\mbox{\boldmath $p$})= \cos \theta_{\mbox{\boldmath $p$}} a^s(\mbox{\boldmath $p$})
		                                         + \sin \theta_{\mbox{\boldmath $p$}}b^{s\dagger }(-\mbox{\boldmath $p$})   .
				                	 \label{eq:8}
\end{equation}
 In contrast,   if $a^s(\mbox{\boldmath $p$})$ and $b^{s\dagger }(-\mbox{\boldmath $p$})$ represent  massive fermion,  they  have  different meaning to the state  because mass is positive definite, and their superposition cannot be  considered.  Hence, this  superposition is  characteristic of massless fermion before symmetry breaking.  

  Let us assign a parameter to each  inertial reference frame in which  antifermions are necessary to protect the temporal order.  The  volume of  such a parameter space is reflected in  $\sin \theta_{\mbox{\boldmath $p$}}$  in Eq.(\ref{eq:8}).  Physically, this $\theta_{\mbox{\boldmath $p$}}$ comes from  the interaction of fermions  with the gauge field, 
but regardless of its details, the following is expected. When $\mbox{\boldmath $p$} =0$,  the difference in  momentum  between 
$ a^s(\mbox{\boldmath $p$})$ and $b^{s\dagger }(-\mbox{\boldmath $p$})$  disappears. Therefore,  the probability  of finding the above inertial frames is maximized,   suggesting  $\cos \theta_{\mbox{\boldmath $p$}}=\sin \theta_{\mbox{\boldmath $p$}}$.  When $\mbox{\boldmath $p$}^2 \rightarrow \infty$,  no such  inertial frame can be found, and $\sin \theta_{\mbox{\boldmath $p$}}  \rightarrow 0$  is expected.  Furthermore, when the coupling $g$ to gauge field disappears,  there is no reason to consider the many-body state, then  $\sin \theta_{\mbox{\boldmath $p$}} =0$ is required.   The same interpretation is possible also for the event at $B$  in Figure.\ref{fig.1}.  
 A new annihilation operator $B^s(-\mbox{\boldmath $p$})$ is defined  as  a superposition of  the annihilation $b^s(-\mbox{\boldmath $p$})$ of massless antifermion  in Figure.\ref{fig.1}(b), and the creation $a^{s\dagger }(\mbox{\boldmath $p$})$  of massless fermion  in 
\ref{fig.1}(a)  
\begin{equation}
		B^s(-\mbox{\boldmath $p$})= \cos \theta_{\mbox{\boldmath $p$}} b^s(-\mbox{\boldmath $p$})
		                                         - \sin \theta_{\mbox{\boldmath $p$}}a^{s\dagger }(\mbox{\boldmath $p$})   .
				                	 \label{eq:9}
\end{equation}
 This $B^s(-\mbox{\boldmath $p$})$ is orthogonal to $A^s(-\mbox{\boldmath $p$})$ \cite{sup}.
 These $A^s(\mbox{\boldmath $p$}')$ and $B^s(-\mbox{\boldmath $p$}')$ describe  not only the two-body state,  but also  general  many-body states.   These  superposed operators   define the  lowest-energy state $|\widetilde{0}\rangle$  by imposing  $A^s(\mbox{\boldmath $p$})|\widetilde{0}\rangle =B^s(-\mbox{\boldmath $p$})|\widetilde{0}\rangle=0$ on it \cite{non}   This  lowest-energy state $|\widetilde{0}\rangle$  should have  a  Lorentz-invariant form,  and  is called  {\it physical vacuum}.  The cyclicity of vacuum is  also expected for  this particular vacuum belonging to  a particular Hilbert space.

\subsection{Physical vacuum}

The explicit form of  $|\widetilde{0}\rangle$ is inferred as follows. In the case of  massless fermions  with velocity close to the velocity of light, 
$ \cos \theta_{\mbox{\boldmath $p$}}  \rightarrow 1$ is required,  and the physical vacuum agrees with  the free vacuum. Therefore   
$|\widetilde{0}\rangle$ includes $ \cos \theta_{\mbox{\boldmath $p$}} |0\rangle$.  Conversely,  in the case of  massless fermions  with small momentum,   various inertial frames  with  large relative velocity $v$  can be set so that it  satisfies $t_2'-t_1'<0$  in Eq.(\ref{eq:7}).  Therefore,  massless  fermions and antifrmions  are required  in $|\widetilde{0}\rangle$,  which leads  to  $b^{s\dagger}(-\mbox{\boldmath $p$}) a^{s\dagger}(\mbox{\boldmath $p$}) |0\rangle$.   
 The simplest possible form of $|\widetilde{0}\rangle$  is a  superpositions of $ \cos \theta_{\mbox{\boldmath $p$}} |0\rangle$ and  $\sin \theta_{\mbox{\boldmath $p$}}b^{s\dagger}(-\mbox{\boldmath $p$}) a^{s\dagger}(\mbox{\boldmath $p$}) |0\rangle$. 
  Such a superposition is possible  for all $\mbox{\boldmath $p$}$, and $|\widetilde{0}\rangle$  is the product of these  superpositions (see Appendx.B) 
\begin{equation}		                                        
	|\widetilde{0}\rangle =\prod_{p,s} \left[ \cos \theta_{\mbox{\boldmath $p$}} 
		          +  \sin \theta_{\mbox{\boldmath $p$}} b^{s\dagger}(-\mbox{\boldmath $p$}) a^{s\dagger}(\mbox{\boldmath $p$}) \right]|0\rangle   .
				                	 \label{eq:135}
\end{equation}
 When the suitable form  for  the  relativistic  many-body states is required, broken-symmetry vacuum manifests itself in the lowest-energy state of  the intermediate states of the  massless Dirac fermions. (The real particle  in  $A^{s\dagger}(\mbox{\boldmath $p$}) |\widetilde{0}\rangle$ or $B^{s\dagger}(\mbox{\boldmath $p$}) |\widetilde{0}\rangle$ is an objective existence common  to all observers, and follows the Lorentz transformation, whereas the massless particle in $|\widetilde{0}\rangle$ is not such an existence. In this sense, the Lorentz transformation which connects different observers  does not apply to the latter, and $|\widetilde{0}\rangle$ is Lorentz invariant.)
(This  $|\widetilde{0}\rangle$ was first introduced to elementary-particle physics by \cite{nam} in analogy with superconductivity \cite{sup}. The above derivation shows that it does not depend on the attractive interaction, but has generality that it depends  only  on the kinematical requirement.)

 \subsection{Pair annihilation of maassless fermion-antifermion pairs  to  gauge boson}
   The broken-symmetry vacuum does not end only  with the creation of fermion and antifermion. Due to  $\bar{\varphi}(i\partial_{\mu}+gB_{\mu})\gamma^{\mu}\varphi $, the  massless fermion-antifermion pair  with opposite momentum in Eq.(\ref{eq:135})  annihilate to a  gauge boson 
with a 4-momentum  $(2p_0, \mbox{\boldmath $0$})$ in the center-of-mass frame of the pair, and  this  gauge  boson annihilates to  other massless  fermion-antifermion   pair in  $|\widetilde{0}\rangle$.    Such a $s$-channel process  between massless objects possesses  no threshold energy, and it results in  an equilibrium state between the massless  fermion-antifermion  pairs and the gauge bosons.  
   At each  point in space,  the total  field-energy of the $U(1)$ gauge field   $B_{\mu}$
\begin{equation}
		\int T^{00}(x)d^3x= \frac{1}{4\pi} \int \left (-F^{0\mu}F_{0\mu}+\frac{1}{4}F^{\mu\nu}F_{\mu\nu} \right ) d^3x  \equiv   \widehat{ \beta},
						                	 \label{eq:14}
\end{equation}
 condenses,  which is a coherent collection of gauge bosons created  at different  points, and  is a  Lorentz- and gauge-invariant scalar quantity  with the dimension of mass.   
To incorporate  such a $\widehat{ \beta}$ to $|\widetilde{0}\rangle$,   the free vacuum  $|0 \rangle$ in the right-hand side of Eq.(\ref{eq:135})  is replaced by a condensed vacuum  $|0_r \rangle$ satisfying  $\langle 0_r | \int T_c^{00}(x)d^3x | 0_r \rangle  \equiv \langle \widehat{ \beta} \rangle  \neq 0 $.  This $|0_r \rangle$ is the lowest-energy  state of the condensed gauge bosons, and its explicit form  is  to be studied in the  future \cite{bos}.
 Here, leaving the explicit form of $|0_r\rangle$ aside,    Eq.(\ref{eq:135}) is redefined as follows
\begin{equation}
		|\widetilde{0}\rangle=\prod_{p,s} \left[ \cos \theta_{\mbox{\boldmath $p$}} 
    +\sin \theta_{\mbox{\boldmath $p$}} e^{i\alpha(x)}b^{s\dagger}(-\mbox{\boldmath $p$}) a^{s\dagger}(\mbox{\boldmath $p$}) \right]|0_r\rangle   .
						                	 \label{eq:16}
\end{equation}
 A  phase factor $\exp[\alpha(x)]$ with respect to  $U(1)$ symmetry  appears   at each point in space-time \cite{cry}.

\section{Dynamical consequences after spontaneous symmetry breaking}
The broken-symmetry vacuum $|\widetilde{0}\rangle$ has a kinematical origin, but produces some  dynamical consequences.   In the Higgs model, symmetry breaking and its  consequence are derived by adding    
\begin{eqnarray}
    L_h(x) &=& |(i\partial_{\mu}+gB_{\mu})(v_h+h)|^2 
                          \nonumber \\
                & -& \mu ^2|v_h+h|^2 -\lambda |v_h+h|^4   
                             \nonumber \\
                 &-& \frac{m_f}{v_h} (v_h+h) \bar{\varphi} \varphi  ,
		 \label{eq:01}
\end{eqnarray}
to $L_0(x)$, where $h=h_1+ih_2$ \cite{eng} \cite{hig}.  In what follows,  we  derive the consequences  without the help  of  $L_h(x)$, and examine the meaning of parameters in $L_h(x)$.

\subsection{Pair production of massless fermions} 
 The condensed field-energy $\widehat{ \beta} $ in Eq.(\ref{eq:14}) is a coherent collection of gauge bosons, which  can create  massless fermion-antifermion pairs.  QED has a good example of this,  where a strong  electric  field $E_{\mu}= F_{0,\mu}$  creates  massive fermion-antifermion pairs.  We know the formula of  pair-production rate $ \Gamma(E\rightarrow f^+f^-)$ 
   by Schwinger \cite{sch}
\begin{equation}
		 \Gamma(E\rightarrow f^+f^-) = \frac{\alpha E^2}{\pi^2} \sum_{n=1}^{\infty} \frac{1}{n^2}  \exp\left( -\frac{n\pi m^2}{\alpha E} \right)    ,
				                	 \label{eq:75}
\end{equation}
where $\alpha=e^2/(4\pi)$.  In QED, strong electric field is necessary to create massive fermion-antifermion pairs, but such a strong field is not required to create massless fermion-antifermion pairs.  Translate this $ \Gamma(E\rightarrow f^+f^-)=2Im(L_{EH})$ ($L_{EH}$ is the Euler-Heisenberg Lagrangian \cite{eul})  into the situation of $L_0(x)$. Replace  the macroscopic field-energy density  $E^2$ in 
Eq.(\ref{eq:75}) with  $T^{00}_c(x)$  of the microscopic field.  In our case,   all energy of the field is put into the  kinetic energy of created massless  particles, and therefore $m^2$
 in Eq.(\ref{eq:75})  is replaced   by $\mbox{\boldmath $p$}^2$.   Incorporate  Eq.(\ref{eq:75}) into the Lagrangian density representing a states process ($\mbox{\boldmath $p$}^2 =0$, that is, $m^2=0$ in Eq.(\ref{eq:75})).   The coupling of the massless fermion and antifermion to the condensed  field-energy  
$\widehat{\beta}$ is as follows
 \begin{equation}
\bar{\varphi}(x)  \left [  \frac{ g^2 }{ 8\pi ^3}  \int T^{00}_c(x)d^3x  \sum^{ \infty}_{n=1}  \frac{1}{n^2}\right]   \varphi(x)
=  \frac{ g^2}{48 \pi } \bar{\varphi}(x)  \widehat{\beta}  \varphi(x) 	,
			                                    \label{eq:76}
\end{equation}                                                            
(here $\sum_{n=1}^{\infty}n^{-2}= \pi^2/6$ is used), which is a simplified  $L_{EH}$ of this case.

 \subsection{Crossing symmetry and fermion's mass}
Crossing symmetry is used  to convert the amplitude by Eq.(\ref{eq:76})  into  that of  a $t$-channel process between fermi fields. There, the  
$\widehat{\beta}$ serves  as a  mean field acting  on  the fermion and antifermion   as follows
\begin{equation}
	\bar{\varphi}(x) \left( i \sla{\partial } + \left[ \widehat{g}^2 \widehat{ \beta} + gB_{\mu} \gamma ^{\mu}  \right]\right) \varphi(x) 	,
			                                    \label{eq:17}
\end{equation}
 where  $ g^2/(48 \pi) = \widehat{g}^2$.  
The physical vacuum $|\widetilde{0}\rangle$ is a stable state with the lowest-energy.    Hence,     Eq.(\ref{eq:17}) sandwiched between $\langle \widetilde{0}|$ and $|\widetilde{0}\rangle$ is diagonal  with respect to  $A^{s\dagger}(\mbox{\boldmath $p$}) A^{s}(\mbox{\boldmath $p$})$ and $B^{s\dagger}(-\mbox{\boldmath $p$}) B^{s}(-\mbox{\boldmath $p$})$ for all $\mbox{\boldmath $p$}$. The mean field   $\widehat{ \beta}$  is  an average of many degrees of freedom, and therefore it does not easily change through the creation  or  annihilation of each  fermion.  Hence,  $\widehat{ \beta}$ is approximated  by a constant vacuum-expectation-value (VEV)  $\langle \widetilde{0}| \widehat{ \beta} |\widetilde{0}\rangle = \langle \widehat{ \beta} \rangle$.  Following the same procedure as in  \cite{sup},   using  the reverse relation of Eqs.(\ref{eq:8}) and (\ref{eq:9}) in  Eq.(\ref{eq:17}),   
the condition for Eq.(\ref{eq:17})  to be diagonal is obtained 
 \begin{eqnarray}
	\cos ^2\theta_{\mbox{\boldmath $p$}} &=&\frac{1}{2}\left(1+\frac{\epsilon_p}{\sqrt{\epsilon_p^2+( \widehat{g}^2  \langle \widehat{ \beta} \rangle )^2} }\right)   ,  
	             \nonumber \\
	\sin ^2\theta_{\mbox{\boldmath $p$}} &=&\frac{1}{2}\left(1-\frac{\epsilon_p}{\sqrt{\epsilon_p^2+( \widehat{g}^2 \langle \widehat{ \beta} \rangle )^2} }\right)   .
						                	 \label{eq:23}
\end{eqnarray}
This condition satisfies  $\cos \theta_{\mbox{\boldmath $p$}}=\sin \theta_{\mbox{\boldmath $p$}}$ at $\mbox{\boldmath $p$}=0$,   and $\sin \theta_{\mbox{\boldmath $p$}} \rightarrow 0$ at $\mbox{\boldmath $p$} \rightarrow \infty$  as expected.  

The diagonalized form of  Eq.(\ref{eq:17})  includes $\sqrt{\epsilon_p^2+( \widehat{g}^2 \langle \widehat{ \beta} \rangle )^2}
\bar{\psi} (p) \psi(p)$, where $\psi(p)$ represents an operator of massive fermion. 
The fermion's mass $m_f$   arises from the condensed  field-energy $\widehat{ \beta}$ of the gauge  field  $B_{\mu}$ 
\begin{equation}
	m_f=  \widehat{g}^2  \langle \widehat{ \beta} \rangle = \frac{g^2}{48\pi} \langle \widetilde{0} | \int T_c^{00}(x)d^3x |\widetilde{0}\rangle	.		
			                                        \label{eq:245}
\end{equation}
 This  mass comes from the fact that massless  antifermions are needed  to constitute  the relativistic  many-body  states  (This situation  parallels the fact in classical  physics that  $E=c\sqrt{p^2+m^2c^2}$, in which the mass $m$ has its origin,   comes from the relativistic  relationship between  energy and momentum.)
This  derivation  does not depend on  whether the effective  interaction between massless fermions  is  attractive or 
 repulsive \cite{sta}.

 \subsection{Goldstone mode and massive gauge boson}
 The physical vacuum $ |\widetilde{0}\rangle$ is not a simple system, and therefore     the response of  $ |\widetilde{0}\rangle$  to $B_{\mu}$ involves  a non-linear relation.  The minimal interaction  $L_0^{min}(x)=\bar{\varphi}(x)(i\partial_{\mu}+gB_{\mu})\gamma^{\mu}\varphi(x) $ itself changes to an effective one  due to  the  perturbation to $|\widetilde{0}\rangle$  by ${\cal H} _I(x)= g j^{\mu}(x) B_{\mu}(x)$. (The perturbation to  $ |\widetilde{0}\rangle$ by $ {\cal H} _I(x)$   does not  double count the effect, because the physical vacuum  $ |\widetilde{0}\rangle$ in Eq.(\ref{eq:16}) does not come from 
$ {\cal H} _I(x)$, but  is selected for the kinematical reason.)  Consider a perturbation expansion of  $\int d^4xL_0^{min}(x)$  in powers of $g$ 
 \begin{eqnarray}
     \lefteqn{ \langle \widetilde{0}|  \int  d^4 x_1 L_0^{min}(x_1)
                                                    exp\left( i\int {\cal H} _I(x_2) d^4x_2 \right) |\widetilde{0}\rangle }
                                                        \nonumber \\
                   &&=   \langle \widetilde{0}|   \int d^4x_1 \bar{\varphi}(x_1) \gamma ^{\mu}[i\partial_{\mu} + gB_{\mu}(x_1)] \varphi(x_1)  |\widetilde{0}\rangle
                                                           \nonumber \\
                     &&+     \langle \widetilde{0}|   \int d^4x_1 L_0^{min}(x_1)
                                                  ig\int d^4x_2 j^{\nu}(x_2) B_{\nu}(x_2)  |\widetilde{0}\rangle   
                                                                                   + \cdots      ,    
                                                                                   \nonumber \\                             
                                                                                               	 \label{eq:34}  
\end{eqnarray}

(1)  In  the last line of  Eq.(\ref{eq:34}),  $B_{\nu}(x_2)$ couples to $\bar{\varphi}(x_1)i\partial^{\mu}\gamma _{\mu} \varphi(x_1)$  in $L^{min}_0(x_1)$.    Integrate  this term partially over $x_1$ in $\bar{\varphi}(x_1)i\partial^{\mu}\gamma _{\mu} \varphi(x_1)$,  and use the fact that  $ \varphi(x_1)$ vanishes at $x_1 \rightarrow \infty$. 
 As a result,   two types of terms  appear, one including $i\partial^{\mu}\bar{\varphi}(x_1)\gamma _{\mu} \varphi(x_1)$,  and  the other including $\partial^{\mu} |\widetilde{0}\rangle$. Because the physical vacuum $ |\widetilde{0}\rangle$  in Eq.(\ref{eq:16})  has an explicit $x$-dependence in the phase  $\alpha(x)$,   the integration in Eq.(\ref{eq:34}) extends to this $\alpha(x)$. The latter term is given by 
 \begin{eqnarray}
	& g& \langle \widetilde{0}|    \int d^4x_1    j_{\mu}(x_1)
	                             \int d^4x_2   j^{\nu}(x_2) B_{\nu}(x_2)   \partial^{\mu} |\widetilde{0}\rangle 
	                                        \nonumber \\
     +& g&  \partial^{\mu} \langle \widetilde{0}|    \int d^4x_1       j_{\mu}(x_1)
	                             \int d^4x_2   j^{\nu}(x_2) B_{\nu}(x_2)    |\widetilde{0}\rangle  .
	                                          \nonumber \\
	                                    						                	 \label{eq:374}
\end{eqnarray}
Here  $\partial^{\mu} |\widetilde{0}\rangle $ is a product of $ \partial^{\mu}\alpha (x)$ and $ \prod_{p,s}\sin \theta_{\mbox{\boldmath $p$}} e^{i\alpha(x)}b^{s\dagger}(-\mbox{\boldmath $p$}) a^{s\dagger}(\mbox{\boldmath $p$}) |0_r\rangle$.  
These $x_1$ and $x_2$ are separated only microscopically  in space-time. If we observe this phenomenon from a far distant point, it would appear to be a local phenomenon at   $X=(x_1+x_2)/2$.  The relative motion along $Y=x_2-x_1$ is indirectly observed as a coefficient appearing in  Eq.(\ref{eq:374}).  For the observer at a distant  space-time point, it is useful to  rewrite   $d^4x_1d^4x_2$ in Eq.(\ref{eq:374}) to $d^4Xd^4Y$. The influence of the physical vacuum $ |\widetilde{0}\rangle$ appears in  Eq.(\ref{eq:374})  through  the following coefficient  for $\mu=\nu$,
 \begin{equation}
	m_B^2=   g^2  \langle \widetilde{0}| \int  j_{\mu}(Y) 
	                                                           j^{\mu}(0) d^4Y  |\widetilde{0}\rangle 	 .	
	                                                           		   	 \label{eq:36}
\end{equation}
 With this Lorentz- and gauge-invariant  $m_B^2$,  Eq.(\ref{eq:374})  is rewritten as 
\begin{equation}
	\frac{2i }{g} m^2_B \int B_{\mu}(X)\partial ^{\mu} \alpha (X) d^4X  \equiv  m_B \int B_{\mu}(X) \partial^{\mu} G(X) d^4X   .
	                              						                	 \label{eq:38}
\end{equation}
Here the Goldstone mode is defined as $G (X)= 2 i g^{-1}m_B \alpha (X)$. (In the Higgs model, the coupling of the Goldstone mode $h_2$ in $h=h_1+i h_2$ to the gauge boson is derived  from the phenomenological term  $ |(i\partial_{\mu}+gB_{\mu})(v_h+h)|^2$. In contrast,   the coupling of $G$ to $B_{\mu}$  in Eq.(\ref{eq:38}) comes from the response of the physical vacuum $ |\widetilde{0}\rangle$ to $B_{\mu}$.)

(2)  In the last line of Eq.(\ref{eq:34}),  $B_{\nu}(x_2)$ also couples to $B_{\nu}(x_1)$ in $L_0^{min}(x_1)$,  yielding the following  two-point-correlation function
\begin{equation}
		  \langle \widetilde{0}|    \int d^4x_1 {\cal H} _I(x_1)     \int d^4x_2  {\cal H} _I(x_2)    |\widetilde{0}\rangle  .
		                  						                	 \label{eq:35}
\end{equation}
The correlation of $ \bar{\varphi}(x_1) \gamma ^{\mu} \varphi (x_1)$ with $ \bar{\varphi}(x_2)\gamma ^{\nu}  \varphi(x_2)$  appears  here when  $\mu=\nu$ as follows     
 \begin{eqnarray}
&g^2& \int \langle \widetilde{0}|    \int  j_{\mu}(x_1)   d^2x_1  
                                                \int   j^{\mu}(x_2)    d^2x_2   |\widetilde{0}\rangle
                                        \nonumber \\        
       &\times&     B^{\mu}(x_1)B_{\mu}(x_2) d^2x_1d^2x_2 
                                                \nonumber \\
=& g^2&  \int   \langle \widetilde{0}|   \int  j_{\mu}(Y) 
                                                  j^{\mu}(0) d^4Y  |\widetilde{0}\rangle \times  B^{\mu}(X)B_{\mu}(X) d^4X                                         
                                                    \nonumber \\
  = &m_B^2&  \int  B^{\mu}(X) B_{\mu}(X) d^4X  .
	 						                	 \label{eq:31.8}                                              	
\end{eqnarray}

(3) In the system without the long-range force, the global phase-rotation  of fermion requires no energy,  and therefore the propagator of the Goldstone mode is given by
\begin{equation}
	\int \frac{dX^4}{(2\pi)^4} \langle \widetilde{0}| T[ G(X) G (0) ] |\widetilde{0}\rangle e^{iqX}= \frac {i} { q^2 }  .
		 \label{eq:381}
\end{equation}
 However, the long-range force mediated by the gauge boson  prohibits  the global free  rotation of the phase $\alpha(x)$, then preventing  the Goldstone mode. This discrepancy is solved  by the generation of a  gauge-boson's mass that converts  the long-range force into a short-range one. 

 The Fourier transform of Eqs.(\ref{eq:38}) and (\ref{eq:31.8})  are given by  $m_Bq^{\mu} G(q) B_{\mu}(q)$ and $m_B^2B^{\mu}(q)B_{\mu}(q)$, respectively. Following the usual way, regard the former as a  perturbation to the latter,  and the second-order perturbation is obtained as
\begin{eqnarray}
	\lefteqn{B^{\mu}(q) \left [im_B^2g^{\mu\nu} - m_Bq^{\mu}  \frac{i}{q^2} m_Bq^{\nu} \right] B_{\nu}(q)	}
	                                    \nonumber \\
	                  &&   = im_B^2 \left( g^{\mu\nu} - \frac{q^{\mu}q^{\nu}}{q^2} \right) B^{\mu}(q)B^{\nu}(q).
		 \label{eq:382}
\end{eqnarray}
Adding this term to the Fourier transform of $-\frac{1}{4} F^{\mu\nu}F_{\mu\nu}$,  and performing an  inverse transformation on  the resulting matrix, we obtain
\begin{eqnarray}
		       D^{\mu\nu}(q) &=& \frac{-i}{q^2-m_B^2} \left( g^{\mu\nu} - \frac{q^{\mu}q^{\nu}}{q^2} \right) 
		                    \nonumber\\
		                         &\equiv& i D(q^2) \left( g^{\mu\nu} - \frac{q^{\mu}q^{\nu}}{q^2} \right),
		 \label{eq:384}
\end{eqnarray}
which is the propagator of the massive gauge boson in the Landau  gauge.

(4)  The  Goldstone mode also  couples to fermions  directly  in the second line of Eq.(\ref{eq:34}). The partial integration  of the zeroth-order term of $B_{\mu}$ over $x_1$ yields  two types of  terms, one including  $i\partial^{\mu}\bar{\varphi}(x_1)\gamma _{\mu} \varphi(x_1)$,  and the other including  $\partial^{\mu} |\widetilde{0}\rangle$. The latter term is given by 
\begin{equation}
		i \langle \widetilde{0}|   \int d^4x_1 j^{\mu}(x_1) \partial_{\mu} |\widetilde{0}\rangle  + i  \partial_{\mu} \langle \widetilde{0}|   \int d^4x_1 j^{\mu}(x_1) |\widetilde{0}\rangle    .
		 \label{eq:385}
\end{equation}
 With  $ \partial_{\mu}\alpha(x)$ in $ \partial_{\mu} |\widetilde{0}\rangle$, and with the definition $G(x)= 2i g^{-1}m_B\alpha(x)$,  
 Eq.(\ref{eq:385}) is rewritten to the coupling of the  Goldstone mode to fermion
\begin{equation}
		 \frac{g}{m_B}  \langle \widetilde{0}|   \int d^4x_1    \bar{\varphi} (x_1)\gamma^{\mu} \varphi (x_1)  \partial_{\mu}G(x_1) |\widetilde{0}\rangle .
		 \label{eq:3851}
\end{equation}
This coupling is different from the corresponding term $g(m_f/m_B)\int d^4x\bar{\varphi} (x) \varphi (x)h_2(x)$ in the Higgs model, because it does not come from the Yukawa coupling. 
In summary,  four  additional terms to $L_0(x)$,  $m_B^2B^{\mu}(q)B_{\mu}(q)$, $m_B B_{\mu}(x)  \partial^{\mu} G (x)$, $( \partial_{\mu}G(x))^2$  and  $gm_B^{-1} \bar{\varphi} (x)\gamma^{\mu} \varphi (x) \partial_{\mu}G(x)$ appear in the   Lagrangian density.

\subsection{Higgs-like  excitation }
\subsubsection{Local excitation propagating in space}

 \begin{figure}
	 	\begin{center}
\includegraphics [scale=0.4]{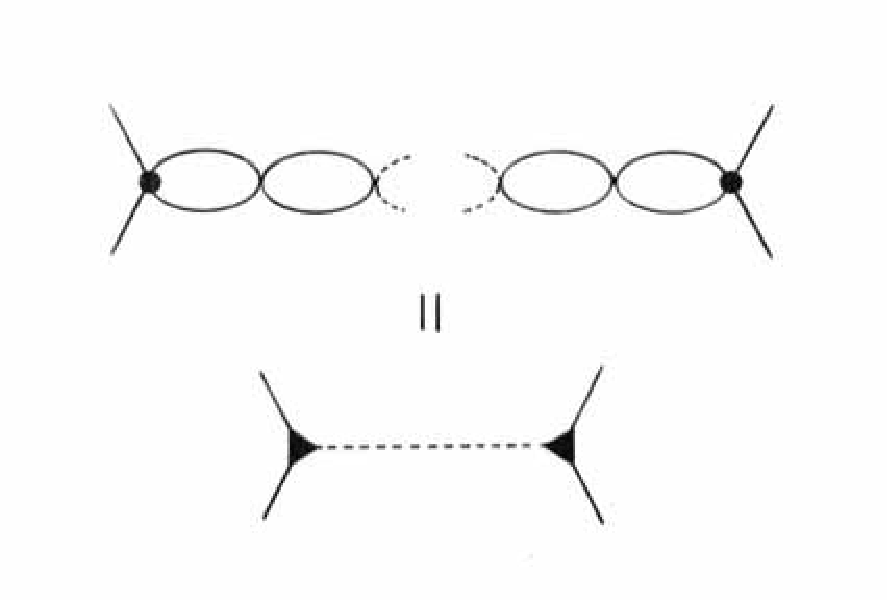}
\caption{Chain of pair creation and annihilation of virtual massless fermion-antifermion pairs}
\label{fig.15}
    	\end{center}
\end{figure}

The  Higgs-like particle found in 2012 is now experimentally examined, and   its properties are  still  hotly debated \cite{atl} \cite{cms}.  Even in this simple
 $L_0(x)$ which  acts on the physical vacuum $|\widetilde{0}\rangle$, the Higgs-like excitation  naturally  appears as a dynamical consequence.
Whenever   the  massless fermions couple  to gauge bosons in  the condensed field-energy $\widehat{\beta}$ in   Eq.(\ref{eq:76}), it is  accompanied by another dynamical process.  The  annihilation of  fermion-antifermion pair  at $x$ in $\widehat{g}^2 \bar{\varphi}(x) \widehat{ \beta} \varphi (x) $ causes  a local excitation  $\beta(x)$ of gauge bosons in  $\widehat{ \beta}$.   This  excitation  $\beta(x)$  annihilate  to other massless fermion-antifermion pair at $x'$.  Hence, it  propagates in space through a chain of  creations and annihilations of massless fermion-antifermion pairs as illustrated in  Figure \ref{fig.3}.  Because this  mode  causes a deviation from  the lowest-energy state,  it can be called  {\it Higgs-like excitation mode} $H_{\beta}(x)$.   Because this  excitation has no special  direction in space,  it can be regarded  as a scalar field.  With $m_f= \widehat{g}^2 \langle \widehat{ \beta}\rangle $,   the dynamical  form of Eq.(\ref{eq:76}) is given by  
\begin{equation}
	  \frac{m_f}{ \langle \widehat{ \beta}\rangle} \bar{\varphi}(x) H_{\beta}(x) \varphi (x)  .
						                	 \label{eq:43}
\end{equation}
This is a non-minimal interaction, because  the gauge transformation occurs  within $H_{\beta}(x)$ (excitation  from $\widehat{ \beta}$ in  Eq.(\ref{eq:14})),  and does  not cause  the  phase rotation of $\varphi (x)$.  Hence, $\gamma ^{\mu}$ matrix is not there. The  bubble diagrams  in  Figure \ref{fig.3}  shows a series of the creation and annihilation of the  fermion-antifermion  pair 
\begin{equation}
	q^2J(q^2)=  \frac{m_f}{\langle \widehat{ \beta}\rangle} \int_{0}^{\Lambda}\frac{d^4p}{(2\pi)^4} tr\left[ \frac{i}{\sla{p}-m_f} \frac{i}{\sla{p}+\sla{q}-m_f} \right]    .
						                	 \label{eq:46}
\end{equation}
This  $J(q^2)$ has following features.

(1)  $J(q^2)$ has a similar form to the vacuum polarization in QED \cite{qed},  but an important difference is that there are no $\gamma ^{\mu}$ or $\gamma ^{\nu}$  in the trace.  

(2) $\Lambda$ in  Eq.(\ref{eq:46}) is not a cutoff for  regularizing  the divergent  integral, but rather an upper end of the energy-momentum  of the excited massless  fermion-antifermion  pairs.  Eq.(\ref{eq:46}) can be calculated  as if $\Lambda$ is such a cutoff for regularization,  but  since  the upper end $\Lambda$  is a dynamical variable,  it is evaluated, not  as $p_E^2$ in the Euclidian space, but as $p^2=(ip_0)^2-\mbox{\boldmath $p$}^2= -p_E^2$  in the Minkowski space. Hence,  after  4-momentum integration,  the square of upper end  appears as $-\Lambda ^2$. 

According to  the ordinary rule, we obtain
\begin{eqnarray}
	J(q^2)&=& \frac{1}{4\pi^2} \frac{m_f}{\langle \widehat{ \beta}\rangle}
	                       	 \int_{0}^{1}dx \left[ x(x-1) +\frac{m_f^2}{q^2}  \right] 
		     \nonumber\\
		& \times& \ln \left(\frac{x(1-x)\Lambda^2+m_f^2}{x(x-1)q^2+m_f^2}\right)   .
	                              						                	 \label{eq:471}
\end{eqnarray}
A peculiar feature of this $J(q^2)$ is that $m_f^2/q^2$ appears  in the integrand.  With this $J(q^2)$,   the propagator of the Higgs-like excitation mode $H_{\beta}(x)$ is given by 
\begin{equation}
 \int \frac{d^4x}{(2\pi)^4}  \langle \widetilde{0}| T[ H_{\beta}(x) H_{\beta}(0) ] | \widetilde{0}\rangle e^{iqx}   =\frac{1}{q^2\left[1- J(q^2)\right]}      .
						                	 \label{eq:48}
\end{equation}

 \begin{figure}
	 	\begin{center}
\includegraphics [scale=0.6]{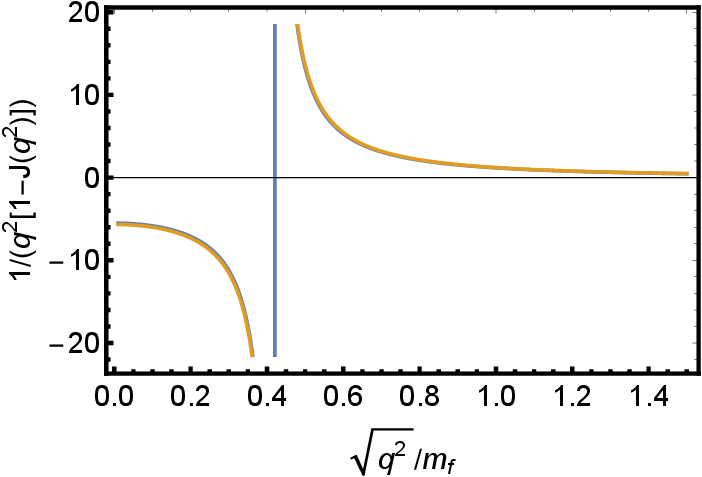}
\caption{Propagator of the Higgs-like collective mode $1/(q^2[1-J(q^2)])$ with  $m_f/ \langle \widehat{ \beta}\rangle=0.64$ and $\Lambda /m_f=800$. }
\label{fig.21}
    	\end{center}
\end{figure}
Figure \ref{fig.21}   shows $1/(q^2[1-J(q^2)])$  when  $m_f/\langle \widehat{ \beta}\rangle=0.64$ and $\Lambda /m_f=800$ are used as an example.  A  pole  appears at $\sqrt{g^2}/m_f= 0.42$  as if it is represented by $1/(q^2-m_H^2) $ with the mass of Higgs-like boson $m_H=0.42m_f$. (The ratio $m_H/m_f$ depends on $\langle \widehat{ \beta}\rangle$ and $\Lambda$.)

  The nature of  non-minimal  interaction in Eq.(\ref{eq:43}) is the reason why  the Higgs-like  excitation mode $H_{\beta}(x)$ has a pole.   The pole structure in Eq.(\ref{eq:48}) arises  from $m_f^2/q^2$ in the integrand of Eq.(\ref{eq:471}),  but this $m_f^2/q^2$  would not appear if 
 $\gamma ^{\mu}$ and $\gamma ^{\nu}$ exist  in the trace of  Eq.(\ref{eq:46}).

In summary, the Higgs-like excitation is described by the following effective  Lagrangian density
\begin{equation}
 (\partial_{\mu} H_{\beta})^2 - m_H^2 H_{\beta}^2  + \frac{m_f}{\langle \widehat{ \beta}\rangle} \bar{\varphi}\varphi H_{\beta} .
						                	 \label{eq:49}
\end{equation}
 The reason why the mass of the Higgs particle has been an unknown parameter in the electroweak model is that it is not a quantity infered from symmetry, but a result of the many-body phenomenon.

\subsubsection{Mass and stability}
\begin{figure}
	 	\begin{center}
\includegraphics [scale=0.65]{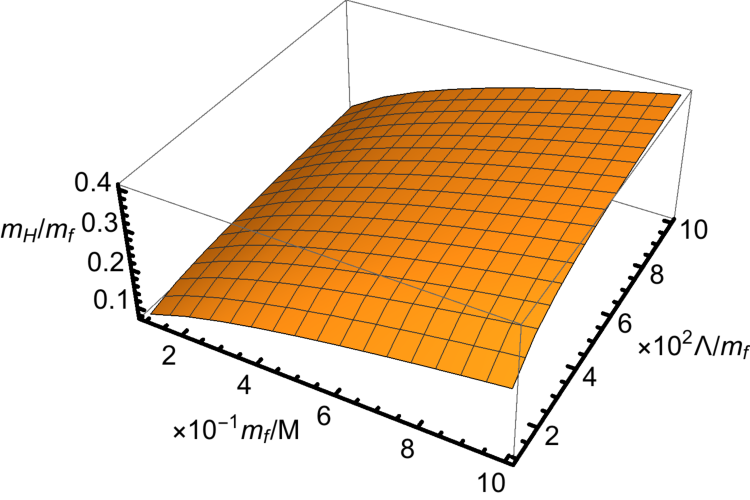}
\caption{The rate of Higgs-like collective mode mass $m_H$ to fermion mass $m_f$ as a function of $m_f/ \langle \widehat{ \beta}\rangle$ and $\Lambda/m_f$.
 ($M$ denotes $ \langle \widehat{ \beta}\rangle$.) }
\label{fig.3}
    	\end{center}
\end{figure}

 The Higgs-like boson's mass $m_H$ is strongly depends on the fermion's mass $m_f$.  But their relationship  also depends on the parameter  $\langle \widehat{ \beta}\rangle$ and  $\Lambda$.  Figure \ref{fig.3} shows $m_H/m_f$,  obtained by solving $J(m_H^2)=1$, as a function of $m_f/\langle \widehat{ \beta}\rangle$ and $\Lambda/m_f$. As these variables increase, it takes much energy $m_H$ to excite the Higgs-like  mode.

This Higgs-like mode is unstable with respect to the decay into  fermion-antifermion  pairs at  $q^2>(2m_f)^2$.  In Eq.(\ref{eq:471}), the  logarithm function  includes $x(x-1)q^2+m_f^2$  in  the denominator, in which $x(x-1)$ is at most $-1/4$ at $x=1/2$. Hence, $x(x-1)+m_f^2/q^2$ becomes negative at $q^2>(2m_f)^2$, which leads to an imaginary energy  in Eq.(\ref{eq:471}). For any fixed $q^2$ at $q^2>(2m_f)^2$, the $x$-value that can contribute to the imaginary energy  in Eq.(\ref{eq:471}) satisfies $x(x-1)q^2+m_f^2<0$, which lies in a region  between the points $x=\frac{1}{2} \pm \frac{1}{2} \delta$, where $\delta = \sqrt{1-4m_f^2/q^2 }$.  Using $Im[-X\pm i\epsilon ]=\pm \pi$, and $y=x-\frac{1}{2}$, we obtain the imaginary part
\begin{eqnarray}
&-& \frac{1}{4\pi^2} \frac{m_f}{\langle \widehat{ \beta}\rangle} (\pm\pi) \int^{(1+\delta)/2}_{(1-\delta)/2} dx \left[x(x-1)+\frac{m_f^2}{q^2} \right]
                                          \nonumber \\
&=&  \pm \frac{1}{4\pi} \frac{m_f}{\langle \widehat{ \beta}\rangle}  \int^{\delta/2}_{-\delta/2} dy \left[ (y^2-\frac{1}{4} )+\frac{m_f^2}{q^2} \right].
						                	 \label{eq:493}
\end{eqnarray}
 Finally, we obtain the propagator of $H_{\beta}(x)$ at $q^2>(2m_f)^2$ as follows 
\begin{eqnarray}
&\int & \frac{d^4x}{(2\pi)^4}  \langle \widetilde{0}| T H_{\beta}(x) H_{\beta}(0) | \widetilde{0}\rangle e^{iqx}	 
                                                            \nonumber \\
       &=& \frac{1}{q^2 - m_H^2 \pm i \displaystyle{ \frac{1}{24\pi} \frac{m_f}{\langle \widehat{ \beta}\rangle}  \sqrt {1-\frac{4m_f^2}{q^2} } (q^2 - 4m_f^2) }}    .
						                	 \label{eq:495}
\end{eqnarray}
in which  $m_H^2$ is used for the real part of the logarithm  function in  Eq.(\ref{eq:471}).
The imaginary part of the self energy increases with increasing $q^2$,  and finally  the excitation  mode becomes unstable.  For the electroweak interaction, however,  we know $m_H=$125 GeV $<2m_t=$346 GeV.  Since $m_H<2m_f$, the structure of the pole-mass around $q^2=m_H^2$ is not affected by the onset of damping  at $q^2>(2m_f)^2$.

\section{Interim summary}
 The total   Lagrangian  density $L(x)$ is as follows. After rewriting $\varphi$ and $\bar{\varphi}$  with $\psi$ and $\bar{\psi}$,  we obtain 
\begin{eqnarray}
    L(x) = &-& \frac{1}{4}F^{\mu\nu}F_{\mu\nu} +  m_B^2 B^{\mu} B_{\mu}   
                                         \nonumber \\
                 &+& \bar{ \psi}(i\partial_{\mu}+gB_{\mu})\gamma^{\mu}\psi - m_f\bar{\psi}\psi 
                                                        \nonumber \\
                      &+& ( \partial_{\mu} G)^2 + m_B B_{\mu} \partial^{\mu} G+  \frac{g}{m_B}  \bar{\psi}  \gamma^{\mu} \psi   \partial_{\mu} G
                                  \nonumber \\
 &+& (\partial_{\mu} H_{\beta})^2 - m_H^2  H_{\beta}^2  + \frac{m_f}{\langle \widehat{ \beta}\rangle} \bar{\psi}\psi H_{\beta}  .
		 \label{eq:50}
\end{eqnarray}

Compared to the Higgs model, this $L(x)$  has the following features.

(1)   The Higgs potential is an economical phenomenology that explains much with a small number of parameters. This is  because it  plays  a double role:  the role of causing symmetry breaking in vacuum and that of  predicting  the  Higgs particle's mass.  Furthermore,   it  stabilizes the broken-symmetry vacuum, and it further  represents  the interaction between the Higgs particle. However, that has made it difficult to imagine the physical processes behind it.  In contrast, such a double role is dissolved  in this $L(x)$. The  broken-symmetry  vacuum  is derived from the kinematical  requirement, and  
the Higgs particle's mass is the result of the many-body phenomenon.  Each  role of the  Higgs potential  is played  by  each physical process.  

(2)  In the Higgs model,   fermion's mass is completely free parameter.  In this $L(x)$,  fermion's mass and fermion's  coupling to Higgs-like excitation are not separated.   Equation.(\ref{eq:17}) represents  a $t$-channel process mediated by $\widehat{ \beta}$, and Eq.(\ref{eq:43}) represents  a $s$-channel process of  the excitation of $\widehat{ \beta}$. 
 
(3) One of important predictions of the Higgs model is that the strength of the Higgs's coupling to fermions is proportional  to the fermion's mass. This is explained by  the $ (m_f/\langle \widehat{ \beta}\rangle) \bar{\psi}\psi H_{\beta} $ in this $L(x)$.  Unlike the Yukawa coupling, however,  the $ (m_f/\langle \widehat{ \beta}\rangle) \bar{\psi}\psi H_{\beta} $ in this $L(x)$ does not lead to the  coupling of  the Goldstone mode $G$  to the fermion.  Rather, such a coupling  $(g/m_B) \bar{\psi}  \gamma^{\mu} \psi   \partial_{\mu} G$ in $L(x)$  arises  from the structure of the physical vacuum.

(4) This $L(x)$  does not  include  $-4\lambda v_h h_1^3 -\lambda h_1^4$ in  the Higgs potential.  Hence, the quadratic divergence does not occur in the perturbation calculation.  The divergence we must renormalize  is only logarithmic one, and  there is no problem of  fine-tuning.

(5) According to the  lattice model,  in which gauge invariance is strictly preserved at each stage of argument,  the VEV  of the gauge-dependent quantity vanishes,   if it is calculated   without gauge fixing, such as  $\langle h(x) \rangle=0 $ for $h(x) \rightarrow h(x)\exp(i\theta (x))$ under $A_{\mu}(x) \rightarrow  A_{\mu}(x) -ig^{-1}\partial_{\mu}\theta(x)$  (Elitzur-De Angelis-De Falco-Guerra theorem) \cite{eli}\cite{gue}.  This is because the local character of gauge symmetry effectively breaks coupling between  degrees of freedom localized in different space-time regions.   If the Higgs particle is an elementary particle, the gauge-fixing dependence of its VEV does not match its fundamental nature.  Instead of the single condensate $v_h=\langle h(x) \rangle$, the present physical  vacuum is characterized by $ \langle \widetilde{0}|  \int  j_{\mu}(Y)  j^{\mu}(0) d^4Y  |\widetilde{0}\rangle$ and $\langle \widetilde{0} | \int T_c^{00}(x)d^3x |\widetilde{0}\rangle = \langle \widehat{ \beta}\rangle$.  The former  determines the gauge-boson's mass $m_B$ in the first-order process of $g$. The latter  determines the fermion's mass $m_f$  in the second-order process of $g$.  Because these two condensates  in the form of integrals are  gauge invariant,  there is no need to worry about  the  vanishing of their VEV as in the VEV of gauge-dependent quantities.


\section{Effective coupling of the Higgs-like  mode}
 For the electroweak interaction,  the Glashaw-Weinberg-Salam (GWS) model  using the  Higgs model is a simple and successful model that does not contradict   almost all experimental results  to date \cite{gla}\cite{wei}\cite{sal}.  So far,  the Higgs coupling to the fermions, especially those  of the third generation,  have been  well parameterized using the Yukawa coupling of fermions to the $h$.  However, for more precise measurements,  there is a possibility of deviation  from such a simple parameterization.
 In the Higgs model,  $\lambda |v_h+h|^4$ in the Higgs potential predicts the triple and quartic self-couplings of the Higgs particle $h_1$. They may correspond to more complex many-body effects  than that  in Figure \ref{fig.3}, which  are well  known  as   collective excitations 
 in the non-relativistic  physics.  The above $L(x)$ predicts some  different results   from those by  the Higgs model.  The next research subject is to  extend it  to the electroweak interaction.

\begin{figure}
	 	\begin{center}
\includegraphics [scale=0.35]{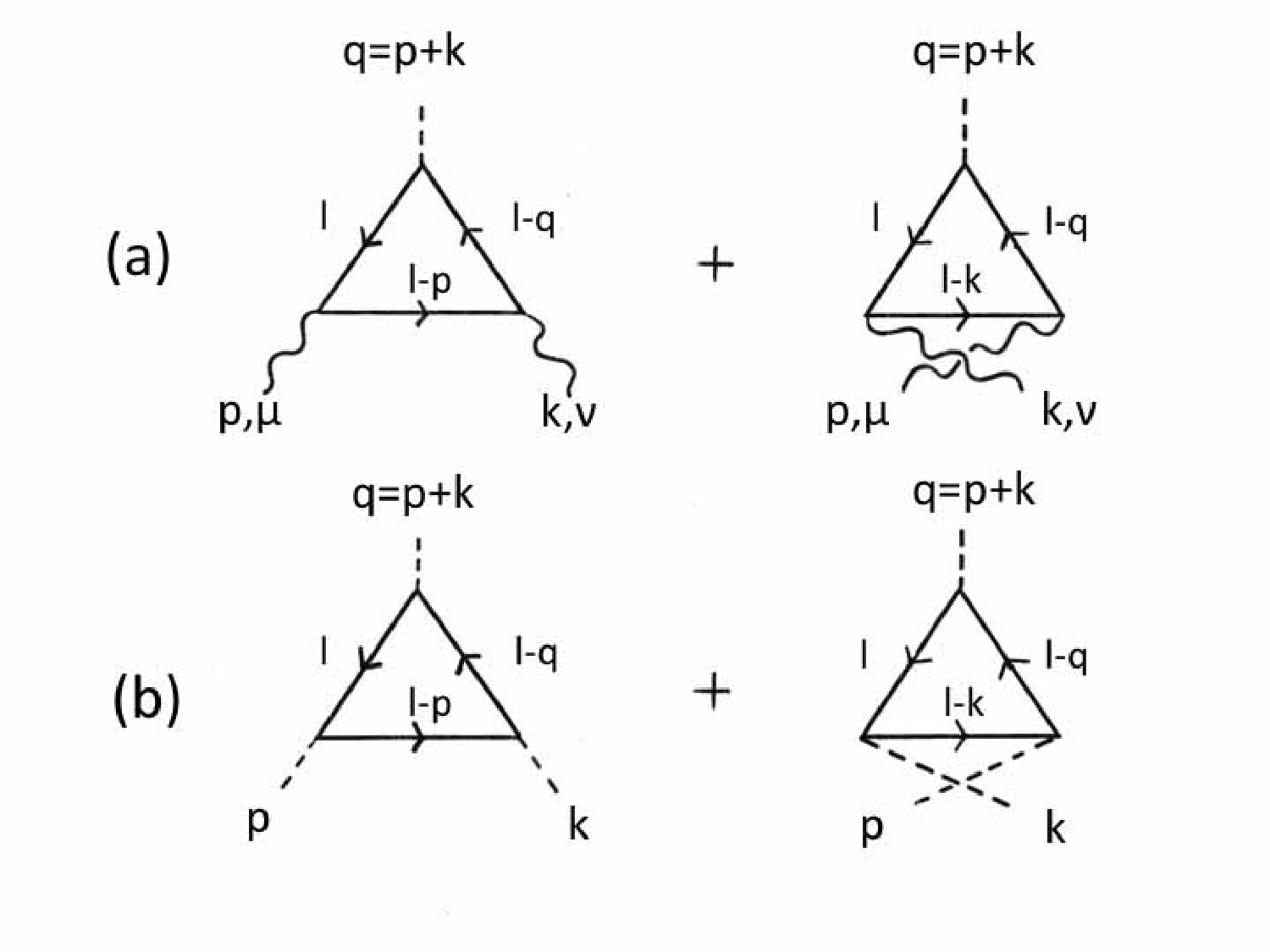}
\caption{ The effective coupling induced by the one-loop diagram  (a) $\Gamma^{\mu\nu} _{BBH}(q, p, k, m_B, m_f) $ in Eq.(\ref{eq:922}), and (b) $F_{H}(q^2, p^2, k^2,  m_f)$ in Eq.(\ref{eq:93}). Wavy and  dotted lines represent $B_{\mu}$ and $H_{\beta}$, respectively.  }
\label{fig.85}
    	\end{center}
\end{figure}

 Recently the decay of the Higgs-like particle to two gauge bosons $W^{+}$ and $W^{-}$ are observed  \cite{atl}\cite{cms}.  
Unlike $ |(i\partial_{\mu}+gB_{\mu})(v_h+h)|^2$ in the Higgs model,  there is no direct coupling of $H_{\beta}$ to $B_{\mu}$ in $L(x)$.  Rather, the effective  coupling of $H_{\beta}$ to $B_{\mu}$  appears first  in the perturbation calculation  through $ (m_f/\langle \widehat{ \beta}\rangle) \bar{\psi}\psi H_{\beta} $ and $g\bar{ \psi}B_{\mu}\gamma^{\mu}\psi$
 comes from one-loop processes  illustrated in Figure \ref{fig.85}.   As a warmup example, we derive such an effective coupling   in the case of  $U(1)$  gauge field, and use it in the calculation of the cross section in Section.7. 
  The result in Section.6 and 7 is obtained using the computing algorithm {\it Package-X} \cite{pat}.
The future precise measurement of the electroweak interaction will determine  whether the deviation from the simple Higgs model exists or not.

\subsection{Coupling to massive gauge bosons}
The effective coupling  term responsible for the $H_{\beta}$ decay into gauge bosons  in Figure \ref{fig.85}(a)  is composed of  $-(m_f/ \langle \widehat{ \beta}\rangle) \bar{\psi}\psi H_{\beta} $ and $ \bar{\psi}gB_{\mu} \gamma^{\mu}\psi$ in  Eq.(\ref{eq:50}).   In coordinate space, such a coupling takes a form 
 \begin{eqnarray}
\lefteqn{g^2 \frac{m_f}{\langle \widehat{ \beta}\rangle} \times   B^{\mu}(x_1)B^{\nu}(x_2) H_{\beta}(x_3) \times  \langle \widetilde{0}|  }
                                         \nonumber \\
&&   T\left [ \int j_{\mu} (x_1) d^4x_1  
                                                \int j _{\nu} (x_2) d^4x_2  \int \bar{\varphi}(x_3)  \varphi(x_3) d^4x_3\right] |\widetilde{0}\rangle . 
                                                                                                               \nonumber \\
	                                                                     	                                                     	 \label{eq:3611}
\end{eqnarray}

(A)  When this coupling is viewed from a distant point in space-time, it looks like a local phenomenon at $X=(x_1+x_2+x_3)/3$ as
 \begin{eqnarray}
\lefteqn{g^2 \frac{m_f}{\langle \widehat{ \beta}\rangle} \int    B^{\mu}(X)B^{\nu}(X) H_{\beta}(X)  d^4X}
                               \nonumber \\
&&\times \langle \widetilde{0}|   T\left [ \int j _{\mu} (x_1)  
                                                \int j _{\nu}(x_2)   \int t(x_3)  d^3x_1 d^3x_1d^3x_3\right] |\widetilde{0}\rangle ,
                                                                                                               \nonumber \\
                                                                               	                                                     	 \label{eq:362}
\end{eqnarray}
where $t(x)=  \bar{\varphi}(x) \sla{\partial } \varphi(x)$.  The coefficient of $B^{\mu}B^{\nu}H_{\beta}$  is a three-point correlation function with a dimension of mass. 
Since the physical vacuum $|\widetilde{0}\rangle$  is  filled with massless fermion-antifermion pairs,   the coefficient of $g_{\mu\nu}B^{\mu}B^{\nu}H_{\beta}$ in Eq.(\ref{eq:362}) has a finite value $V_h$  
 \begin{equation}
	V_h=  \langle \widetilde{0}|   T\left [ \int j _{\mu} (x_1)  
                                                \int j^{\mu} (x_2)   \int t (x_3)  d^3x_1 d^3x_1d^3x_3\right] |\widetilde{0}\rangle,
	  	  						                	 \label{eq:363}
\end{equation}  
 because of  $\prod_{p,s}\sin \theta_{\mbox{\boldmath $p$}} e^{i\alpha(x)}b^{s\dagger}(-\mbox{\boldmath $p$}) a^{s\dagger}(\mbox{\boldmath $p$}) |0_r\rangle $  in $ |\widetilde{0}\rangle$. 
 This $V_h$ plays the role of $v_h$ in  Eq.(\ref{eq:01}). 

(B)  When this coupling is viewed  in high resolution, anomalous momentum-dependent coupling is observed.
  We consider an amplitude $ {\cal M} [ H_{\beta}(p+k)\rightarrow B_{\mu}(p)B_{\nu}(k)] $ in Figure \ref{fig.85}(a) 
\begin{eqnarray}
	\lefteqn{g^2 i {\cal M} [ H_{\beta}(p+k)\rightarrow B_{\mu}(p)B_{\nu}(k)] }    \nonumber \\
	&& \propto  (-ig)^2(-i\frac{m_f}{\langle \widehat{ \beta}\rangle}) 	
	                                                \nonumber \\	
	                   && \quad  \times        \int\frac{d^4l}{(2\pi)^4} tr\left[\gamma^{\mu}\frac{i}{\sla{l}-\sla{p}-m_f} \gamma^{\nu}\frac{i}{\sla{l}-\sla{q}-m_f}\frac{i}{\sla{l}-m_f}\right]   
	                                                  \nonumber \\
	                                     	   && \quad  +  (p \leftrightarrow k,  \mu \leftrightarrow \nu)    .  
		                                              						                	 \label{eq:51}
\end{eqnarray}
Following the standard procedure,  the analytic result is obtained as follows.  
The general form of the interaction between $B_{\mu}$ and  $H_{\beta}$   
 \begin{eqnarray}
	&g^2&  (\frac{m_f}{\langle \widehat{ \beta}\rangle}) \left[ V_hg^{\mu\nu}+m_f\Gamma^{\mu\nu} _{BBH}(q, p, k, m_B, m_f) \right]
	                       \nonumber \\
	&\times & B_{\mu}(p)B_{\nu}(k) H_{\beta}(p+k) ,
	  	  						                	 \label{eq:922}
\end{eqnarray}  
has the following  anomalous momentum-dependence
\begin{eqnarray}
	\lefteqn{ \Gamma^{\mu\nu} _{BBH}(q, p, k, m_B, m_f) } 
	                            \nonumber \\
	    && =   F_1(q^2,p^2,k^2,m_f)g^{\mu\nu} 
	                            \nonumber \\
	 &&+   F_2(q^2,p^2,k^2,m_f) \frac{p^{\nu}k^{\mu}+p^{\mu}k^{\nu}}{m_B^2} 
	                                     \nonumber \\
	                  &&  +  F_3(q^2,p^2,k^2,m_f) \frac{p^{\nu}p^{\mu}+k^{\mu}k^{\nu}}{m_B^2}    .
	                 	                              				                	 \label{eq:925}
\end{eqnarray}
In these $F_i$, the divergences coming from the triangle-loop integral are cancelled to each other in  Eq.(\ref{eq:51}). 

(a) We obtain the leading anomalous   couplings  $ F_1(q^2,p^2,k^2,m_f)$ in  Eq.(\ref{eq:925})  \cite{shi} 
\begin{eqnarray}
	 \lefteqn{F_1(q^2,p^2,k^2, m_f) = }
	                                          \nonumber \\
	 && \frac{8}{(4\pi)^2} \left[ 1+  \frac{(p^2+k^2)q^2 -8p^2k^2}{(q^2-4p^2) (q^2-4k^2) }  f \left(\frac{q}{2m_f} \right) \right] 
	                                           \nonumber \\
	  && -  \frac{8}{(4\pi)^2}  \left[ \frac{p^2}{q^2-4p^2}  f  \left(\frac{p}{2m_f} \right)  + (p \leftrightarrow k)\right]
	                                                     \nonumber \\
	&& -  \frac{2}{(4\pi)^2}    \frac{ q^4 -(6p^2+4m_f^2)q^2 +4p^4 +16p^2m_f^2  }{q^2-4p^2} 
	                                                \nonumber \\
	&& \quad \times  C_0(p^2, p^2, q^2, m_f^2, m_f^2, m_f^2)  
	                                              \nonumber \\
	 &&  - (p \leftrightarrow k),
	                         						                	 \label{eq:565}
\end{eqnarray}
where
\begin{figure}
	 	\begin{center}
\includegraphics [scale=0.65]{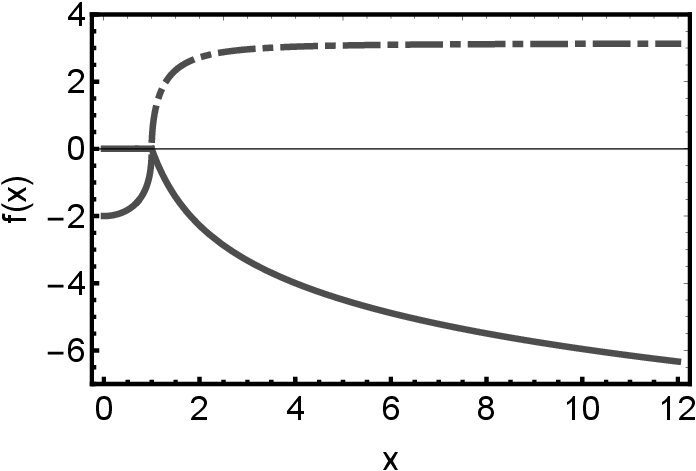}
\caption{Schematic view of $f(x)$. $Re f(x)$ and $Im f(x)$ are represented by a solid, and a one-point-dotted  curve, respectively. }
\label{fig.945}
    	\end{center}
\end{figure}
\begin{figure}
	 	\begin{center}
\includegraphics [scale=0.65]{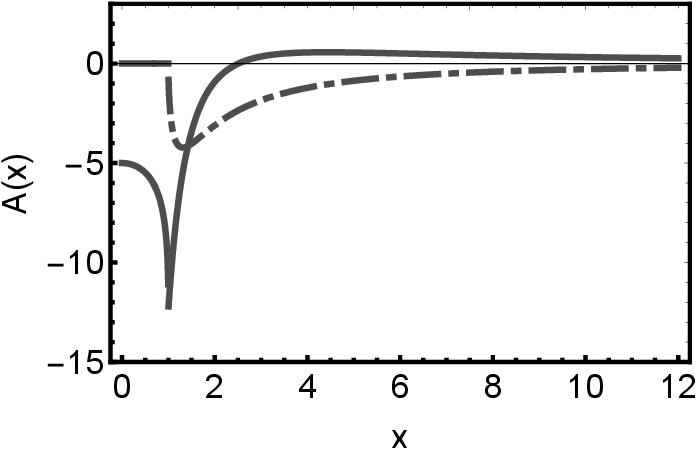}
\caption{Schematic view of $A(x)$. $Re A(x)$ and $Im A(x)$ are represented by a solid, and a one-point-dotted  curve, respectively. }
\label{fig.946}
    	\end{center}
\end{figure}

(1) The function $f(x)$
\begin{equation}
	 f(x)=  \sqrt{1-\frac{1}{x^2}} \ln \left[ 1-2x^2+2x^2 \sqrt{1-\frac{1}{x^2}} \right]  ,
	  						                	 \label{eq:571}
\end{equation}
is illustrated in  Figure.\ref{fig.945}.

(2) $C_0(p^2, p^2, q^2, m_f^2, m_f^2, m_f^2) $  is the scalar $C_0$ function in the Passarino-Veltman integrals \cite{pas}\cite{bar}, 
\begin{eqnarray}
	\lefteqn{C_0(p^2, p^2, q^2, m_f^2, m_f^2, m_f^2) }  
	                                        \nonumber \\
	                          && = \frac{2}{q^2} A(\frac{q}{2m_f})+  \left( \frac{1}{m_f^2q^2} +  \frac{1}{q^4}  A(\frac{q}{2m_f}) \right) p^2 + \cdots ,
	                                         \nonumber \\
	  						                	 \label{eq:5672}
\end{eqnarray}
where
\begin{equation}
	 A(x)=  - \arcsin ^2 x
	  						                	 \label{eq:56741}
\end{equation}
for for $x<1$, and 
\begin{equation}
	 A(x)=   \frac{1}{4} \left(\ln  \displaystyle{ \left[ 1-2x^2+2x^2 \sqrt{1-\frac{1}{x^2}} \right]  }-i\pi \right)^2, 
	 	  						                	 \label{eq:56742}
\end{equation}
for $x>1$,   which is illustrated in Figure.\ref{fig.946}.  Since $f(x) \rightarrow -2$ and $ \arcsin x \rightarrow x$ at $x\rightarrow 0$, $F_1(0,0,0,m_f)=0$ is obtained in Eq.(\ref{eq:565}).

  \begin{figure}
	 	\begin{center}
\includegraphics [scale=0.65]{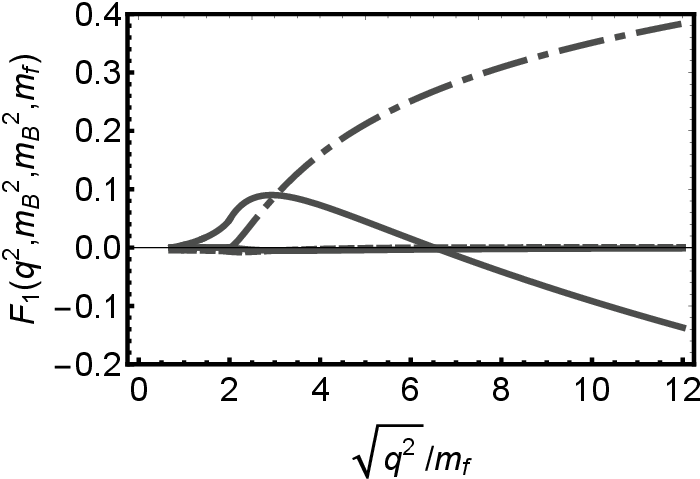}
\caption{ The overall $q^2$-dependence  of the anomalous effective  coupling  $ F_1(q^2,m_B^2, m_B^2,m_f)$; the real part $ReF_1$ (solid curve), and its imaginary part $ImF_1$  (one-point-dotted curve).  The mass of  the gauge boson $B_{\mu}$ is $m_B=80$ GeV. The mass $m_H$ of $H_{\beta}$ and the mass $m_f$ of the fermion in the triangle loop are assumed as  $m_H=123$ GeV and  $m_f=160 $ GeV, respectively.  The condition of $q^2=m_H^2$ corresponds to  $\sqrt{q^2}/m_f=0.76$, from which $F_1(q^2)$ begins to exist.  The solid curve at $\sqrt{q^2}/m_f<2$ is magnified in Figure.\ref{fig.9}. }
\label{fig.95}
    	\end{center}
\end{figure}

 \begin{figure}
	 	\begin{center}
\includegraphics [scale=0.65]{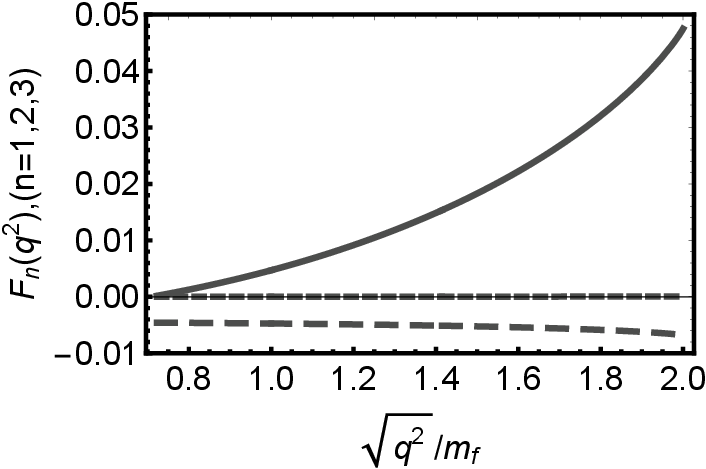}
\caption{Three types of the real part of anomalous effective  coupling at $\sqrt{q^2}<2m_f$, $ F_1(q^2,m_B^2, m_B^2,m_f)$ (solid curve), $ F_2(q^2,m_B^2,m_B^2,m_f) $ (dotted curve), and  $ F_3(q^2,m_B^2,m_B^2,m_f) $  (short dotted curve),  under the same condition as Figure \ref{fig.95}. }
\label{fig.9}
    	\end{center}
\end{figure}

(b) For  the decay $ H_{\beta}(p+k)\rightarrow B_{\mu}(p)B_{\nu}(k)$  illustrated in Figure \ref{fig.85}(a), we numerically calculate  the anomalous effective coupling  at $p^2=k^2=m_B^2$.   Figure \ref{fig.95} shows the overall $q^2$-dependence of  $F_1(q^2,m_B^2, m_B^2,m_f)$: its real part $Re F_1$ (solid curve), and its imaginary part $Im F_1$ (one-point-dotted curve), in which   $m_B=80$ GeV,  $m_H=123$ GeV, and  $m_f=160 $ GeV are used.   The real part  $ Re F_1(q^2,m_B^2, m_B^2,m_f)$ increases with  increasing $\sqrt{q^2}$ to $2m_f$,  and  decreases at $\sqrt{q^2}>2m_f$, then changing its sign. The amplitude of imaginary part $Im F_1(q^2,m_B^2, m_B^2,m_f)$ remains zero  at $\sqrt{q^2}<2m_f$, but gradually increases at  $\sqrt{q^2}>2m_f$.
 
(c)  In the present model, other  effective anomalous couplings $F_2(q^2, p^2,k^2, m_f)$ and $F_3(q^2, p^2,k^2, m_f)$   inevitably appears. 
 (See Appendix).   Figure \ref{fig.9} shows  $Re F_1$, $Re F_2$ and $Re F_3$ at $\sqrt{q^2}<2m_f$,  in the case of same $m_f$, $m_B$,  $m_H $ as in Figure \ref{fig.95}.     $Re F_2$  and   $Re F_3$ are much smaller than  $Re F_1$.   The imaginary parts $Im F_1$,  $Im F_2$ and $Im F_3$ are zero in $\sqrt{q^2}< 2m_f$.

\subsection{Self coupling}
  The effective self-coupling of the Higgs-like  mode  is created by the one-loop process  illustrated  in Figure \ref{fig.85}(b). In coordinate space, this coupling is expressed by another three-point correlation $W_h$  
 \begin{equation}
      W_h=  \langle \widetilde{0}|   T\left [ \int  \widetilde{t} (x_1)  
                                                \int \widetilde{t} (x_2)   \int t (x_3)  d^3x_1 d^3x_1d^3x_3\right] |\widetilde{0}\rangle,
	  	  						                	 \label{eq:36314}
\end{equation}  
where $\widetilde{t}(x)= \bar{\varphi}(x)  \varphi(x)$.    The anomalous momentum-dependent self-coupling   is obtained by the following amplitude
\begin{eqnarray}
	\lefteqn{  i {\cal M} [H_{\beta}(p+k)\rightarrow H_{\beta}(p) H_{\beta}(k)] }    \nonumber \\
	&& \propto (-i\frac{m_f}{\langle \widehat{ \beta}\rangle})^3 	                                              	
	                          \int\frac{d^4l}{(2\pi)^4} tr\left[\frac{i}{\sla{l}-\sla{p}-m_f} \frac{i}{\sla{l}-\sla{q}-m_f}\frac{i}{\sla{l}-m_f}\right]   
	                                      \nonumber \\
	   &&  +  (p \leftrightarrow k)    .                                      
						                	 \label{eq:58}
\end{eqnarray}
\begin{figure}
	 	\begin{center}
\includegraphics [scale=0.65]{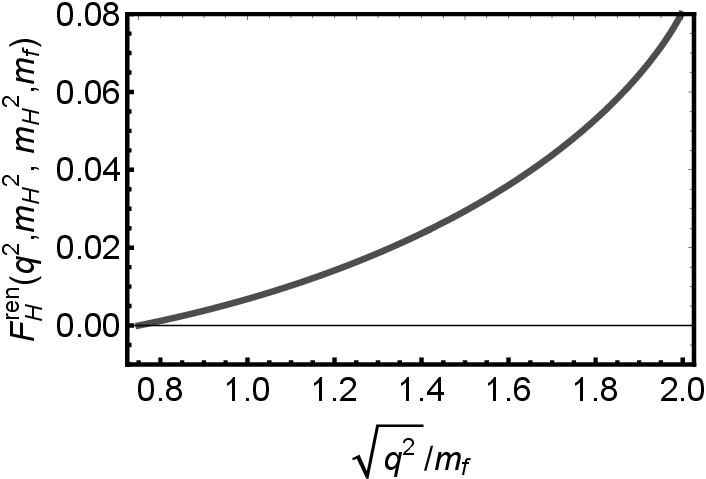}
\caption{The renormalized anomalous effective self-coupling $ F^{ren}_{H}(q^2, m_H^2,m_H^2, m_f)$ of Higgs-like  mode,   for the decay of  off-shell such a mode into  two on-shell such modes,  in the case of same $m_H$ and  $m_f $ as in Figure \ref{fig.95}.  The condition of  $q^2=m_H^2$ corresponds to $\sqrt{q^2}/m_f=0.76$.}
\label{fig.10}
    	\end{center}
\end{figure}
Hence, the effective self-coupling of the Higgs-like  mode has a form such as
 \begin{equation}
	 (\frac{m_f}{\langle \widehat{ \beta}\rangle})^3 [W_h+ m_f F_{H}(q^2, p^2, k^2, m_f) ] H_{\beta}(p) H_{\beta}(k) H_{\beta}(p+k)  ,
	  	  						                	 \label{eq:93}
\end{equation}   
 which corresponds to  $ \lambda v_h$ in Eq.(\ref{eq:01}). This $ (m_f/ \langle \widehat{ \beta}\rangle)^3  W_h$ no longer plays the role of stabilizing the symmetry-broken vacuum as in the Higgs potential in $L_h(x)$.

Following the standard procedure,  we calculate the anomalous self-coupling, and  find that  the divergence $1/\epsilon$ coming from the triangular-loop integrals does  not cancel each other in Eq.(\ref{eq:58}).   We set a condition that $F_{H}(q^2, p^2,k^2, m_f)$ is zero at the on-shell level, and  has a finite value only at the off-shell level $q^2>m_H^2$. Hence, $F_{H}(q^2, p^2, k^2,  m_f)$  must vanish at $q^2=m_H^2$.  (For the divergence that is independent of momentum, it is to be renormalized  to $W_h$.)  We define  the renormalized one as $F^{ren}_{H}(q^2, p^2,k^2, m_f) = F_{H}(q^2, p^2,k^2, m_f)- F_{H}(m_H^2, p^2,k^2, m_f)$, and obtain 
\begin{eqnarray}
	\lefteqn{ F^{ren}_{H}(q^2, p^2,k^2, m_f)
	  =  \frac{8 }{(4\pi)^2} \left[ f \left(\frac{q}{2m_f}\right)- f \left(\frac{m_H}{2m_f}\right) \right]  }
	                             \nonumber \\   
	    &&+ \frac{2 }{(4\pi)^2}  (8m_f^2 - 2p^2 - q^2)  C_0(p^2, p^2, q^2, m_f^2, m_f^2, m_f^2) 
	                                \nonumber \\	                          
	  &&-   \frac{2 }{(4\pi)^2}  (8m_f^2 - 2p^2 - m_H^2)  C_0(p^2, p^2, m_H^2, m_f^2, m_f^2, m_f^2) 
	                               \nonumber \\
	  && +(p \leftrightarrow k)  .
	                               	                          		                						                	 \label{eq:61}
\end{eqnarray}

 For the decay $H_{\beta}(p+k)\rightarrow H_{\beta}(p) H_{\beta}(k)$   in  Figure \ref{fig.85}(b), we numerically calculate  $ F^{ren}_{H}(q^2,  m_H^2,m_H^2, m_f)$.  Figure \ref{fig.10} shows the result under the same $m_H$ and  $m_f $ as in Figure \ref{fig.95}.  This effective coupling gradually increases from zero with  increasing $\sqrt{q^2}$ to $2m_f$.  The overall $q^2$-dependence of $F^{ren}_{H}(q^2, m_H^2,m_H^2, m_f)$  is quite similar to $F_1(q^2, m_B^2,m_B^2, m_f)$, except for its absolute value.


\section{Pair annihilation of fermion and antifermion  to gauge boson pair}

\begin{figure}
	 	\begin{center}
\includegraphics [scale=0.5]{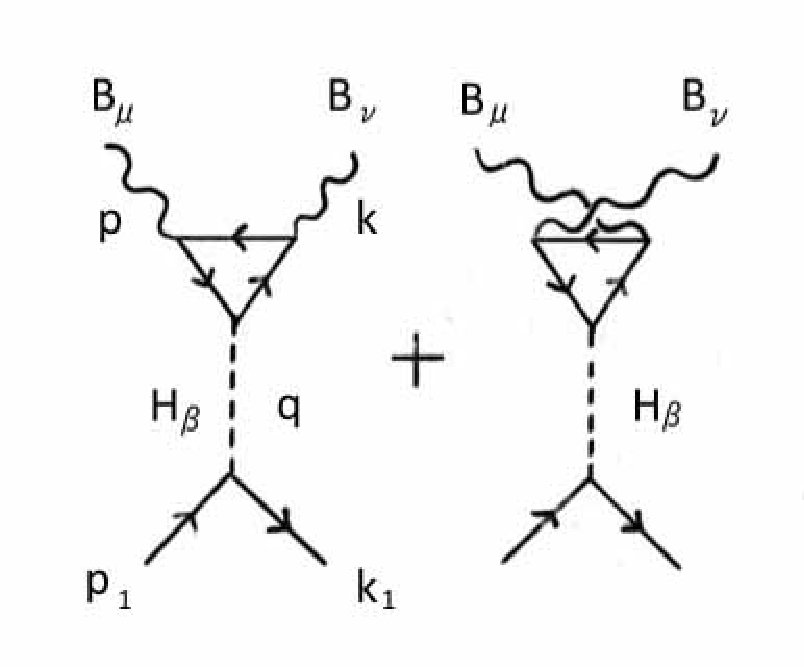}
\caption{ Fermion-antifermion pair annihilation  through the intermediate Higgs-like  mode  to gauge boson pair. }
\label{fig.104}
    	\end{center}
\end{figure}

As an example of the reaction including the production and decay of the Higgs-like  mode,  we consider a $s$-channel process: fermion  $\psi(p_1)$ +  antifermion  $\bar{\psi}(k_1)$ $\rightarrow H_{\beta}(q)$  $ \rightarrow B_{\mu} (p)$ +$ B_{\nu}(k)$  illustrated in Figure.\ref{fig.104}.  This reaction  occurs together with the background reactions, such as the tree process through  the $t$- and  $u$-channel exchanges of fermion. The  cross section of such a tree process is a slowly and monotonously decreasing  function of the center-of-mass energy $E_{cm}$. In the total cross section,  the effective coupling  of the Higgs-like mode  will appear above this almost constant  background cross section.   In view of Figure \ref{fig.95} and \ref{fig.9}, we use only  $F_1(q^2,m_B^2,m_B^2,m_f)$ as the first approximation of the total effective coupling. We obtain the amplitude of  the process in  Figure.\ref{fig.104}  
\begin{eqnarray}
 \lefteqn{ i {\cal M}= \frac{m_f}{\langle \widehat{ \beta}\rangle} \bar{V}(k_1)U(p_1)
        g^2\frac{m_f}{\langle \widehat{ \beta}\rangle}}
                                        \nonumber \\
      && \times \frac{ \displaystyle{ \left[V_h+m_f[ReF_1(q^2)+iImF_1(q^2)]\right] }}{q^2 - m_H^2\pm i \displaystyle{ \frac{1}{24\pi} \frac{m_f}{\langle \widehat{ \beta}\rangle}  \sqrt {1-\frac{4m_f^2}{q^2} } (q^2-4m_f^2) }} 
                         \epsilon^{\mu} (p)\epsilon_{\mu} (k)  ,
                                       \nonumber \\
						                	 \label{eq:611}
\end{eqnarray}
where $ \epsilon^{\mu} (p)$ is one of  polarization vectors of the massive gauge boson,  satisfying $ \epsilon^{\mu} (p)\epsilon_{\mu} (p)=-1$ for longitudinal and transverse polarizations.

Since the incident fermion and antifermion  are lighter than the particles in the triangle loop,  the former are assumed to be massless.  Using $ \sum_{s} U(p_1)\bar{U}(p_1) = \sla{p_1}$ and $\sum_{s} V(k_1)\bar{V}(k_1) = \sla{k_1}$, we obtain  the squared amplitude 
 \begin{eqnarray}
\lefteqn{\frac{1}{4} \sum_{s,s'} | {\cal M} |^2 = g^4 \left(\frac{m_f}{\langle \widehat{ \beta}\rangle}\right)^4  12 (p_1\cdot k_1)   }	
				                                 \nonumber \\
		 && \times  \frac{\displaystyle{ [V_h+ m_f ReF_1(q)]^2+ m_f^2 [ImF_1 (q)]^2}}
    {(q^2 - m_H^2)^2 + \displaystyle{ \left(\frac{1}{24\pi} \frac{m_f}{\langle \widehat{ \beta}\rangle}\right)^2  \left(1-\frac{4m_f^2}{q^2} \right) (q^2-4m_f^2)^2 }}             .
		                                         \nonumber \\
		 \label{eq:615}
\end{eqnarray}

\begin{figure}
	 	\begin{center}
\includegraphics [scale=0.65]{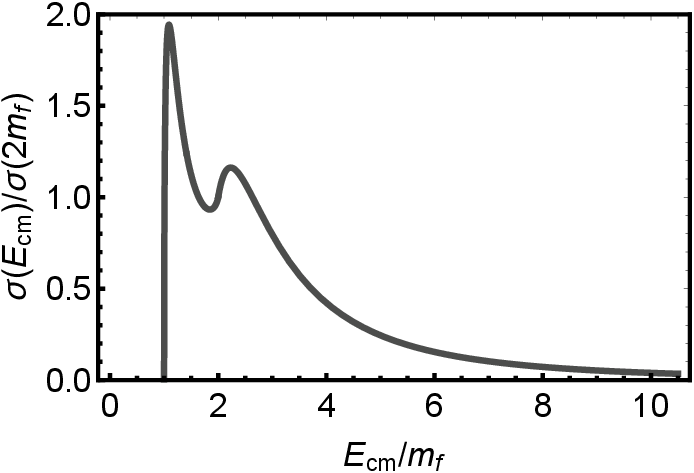}
\caption{The rate of total cross section $\sigma(E_{cm})$ to $\sigma(2m_f)$  in the  fermion-antifermion annihilation to gauge boson pair through the intermediate Higgs-like collective mode,  in the case of $m_B=80$ GeV,  $m_f=160$ GeV, $m_H=123$ GeV and $V_h/m_f= 0.02$. }
\label{fig.105}
    	\end{center}
\end{figure}

In the center-of-mass frame where $p_1=(E_{cm}/2, \mbox{\boldmath $p$}_1 )$ and $k_1=(E_{cm}/2, -\mbox{\boldmath $p$}_1 )$,  we can use $q^2=(p_1+k_1)^2=E_{cm}^2$ and 
$p_1\cdot k_1=E_{cm}^2/2$.  The total cross section $\sigma (E_{cm})$ is given by
\begin{equation}
                  \sigma (E_{cm})= \frac{\pi}{E_{cm}^2}\sqrt{1- \frac{4m_B^2}{E_{cm}^2}} \sum_{s,s'}  | {\cal M} |^2  .
                   						                	 \label{eq:616}
\end{equation}

Figure \ref{fig.105} shows the rate of  total cross section $\sigma (E_{cm})$ to $\sigma (2m_f)$   in the case of  $V_h/m_f= 0.02$, and same $m_f$, $m_B$,  $m_H $ as in Figure \ref{fig.95}. The steep rise of $\sigma$ occurs at $E_{cm}=2m_B$. In contrast to  the Higgs model, $\sigma$ gradually increases below $E_{cm}= 2.3m_f$, and later decreases  at $E_{cm}> 2.3m_f$, which are peculiar feature of the Higgs-like excitation mode. The smaller the rate  $V_h/m_f$ is  in Eq.(\ref{eq:615}),   this gradual increase and decrease  of $ \sigma (E_{cm})$ is is more clearly observed around $E_{cm}=2.3m_f$.
(If $m_H>2m_B$, a  sharp resonance  peak  representing a pole at $q^2=m_H^2$ also appears.) This $E_{cm}$-dependence mainly comes from $ReF_1$ and $ImF_1$. It is not affected by the damping of the Higgs-like  mode $H_{\beta}$, because the effect of damping is weakened by the small factor $(1/24\pi)^2$  in Eq.(\ref{eq:615}). 
  {\it The gradual increase and decrease of the cross section around $E_{cm}=2.3 m_f$ is a key feature for the experimental confirmation  of the physical vacuum}, which is absent in $L_h(x)$.


\section{Renormalizability}

In  the Higgs model, symmetry breaking is caused simply by  changing the sign of the coefficient $\mu ^2$  in the Higgs potential $-\mu^2 |v_h+h|^2+ \lambda |v_h+h|^4$. 
Renormalizability of the Higgs model is systematically proved using  the generating functional with this Higgs potential \cite{abe}. 
(1)  In the symmetric vacuum,  this generating functional is expanded in powers of the Higgs field $h(x)$.  In  the symmetry-broken vacuum,  it is expanded in powers of $h(x)-v_h$.  The algebraic relationship between these two types of expansion can be obtained, which assures that the renormalizability of the theory in the symmetric vacuum  can be transferred to that in the symmetry-broken one. (2) When the Bogoliubov-Parasuik-Hepp-Zimmermann (BPHZ) method is applied to the Ward-Takahashi identities derived from this generating functional,  renormalizability  is systematically proved without reference to the symmetric model \cite{lee}.

In the present model using $L_3(x)$ in Eq.(\ref{eq:50}), however, the Higgs potential is not assumed. Hence, we can not immediately apply this systematic method. Rather,  we  go back to the inductive proof of  renormalizability originated by Dyson in QED \cite{dys}.  Let us consider the following Green's functions:  $S(p)$ for the massive fermion $\psi$,  
\begin{equation}
	  [ S(p) ] ^{-1} = - \gamma \cdot p +m_f - \Sigma ^* (p)   ,
	  						                	 \label{eq:62}
\end{equation}
$D(p^2)$  for the massive gauge boson $B_{\mu}$  in Eq.(\ref{eq:384}),
\begin{equation}
	  [ D (p^2) ] ^{-1} = -p^2 + m_B^2 - \Pi  ^*(p^2)    ,
	  	  						                	 \label{eq:63}
\end{equation}
and $G(p^2)$ for the Higgs-like  mode $H_{\beta}$
\begin{equation}
	 [ G (p^2) ] ^{-1} = -p^2 +m_H^2 - \Pi_H  ^*(p^2)    .
	 	  						                	 \label{eq:64}
\end{equation}
 \begin{figure}
	 	\begin{center}
\includegraphics [scale=0.34]{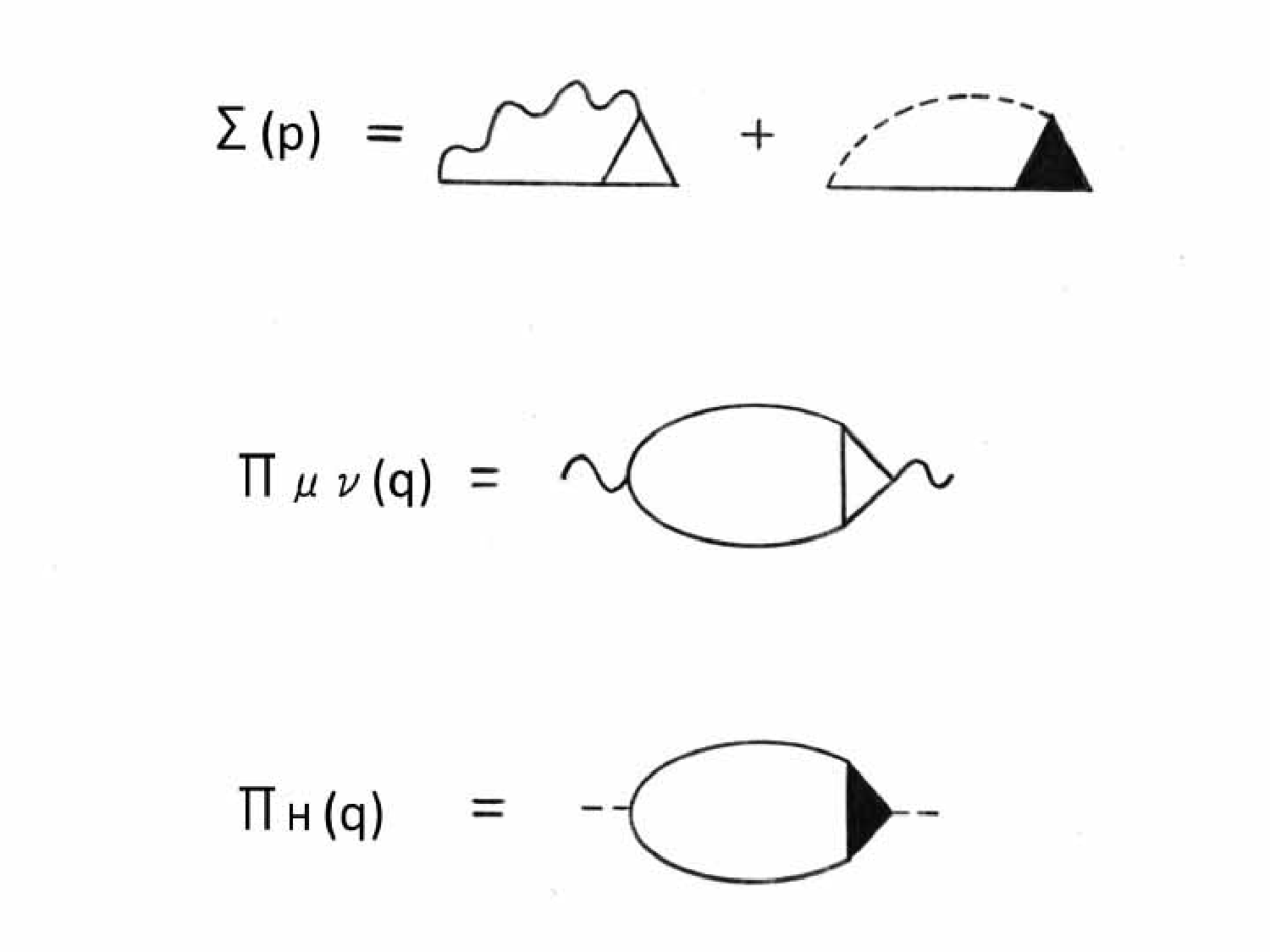}
\caption{(a) Self energy $\Sigma (p)$ of the fermion $\psi$,  (b) Vacuum polarization $\Pi_{\mu\nu}(q)$ of the gauge field $B_{\mu}$, (c) Vacuum polarization 
$\Pi_H(q)$ of the Higgs-like  mode $H_{\beta}$. Wavy and  dotted lines represent $B_{\mu}$ and $H_{\beta}$, respectively. A white  triangle denotes a vertex $\Gamma _{\mu}$ between $B_{\mu}$ and $\psi$,  and a black triangle denotes $\widehat{\Gamma}_H$  between $H_{\beta}$ and $\psi$. }
\label{fig.5}
    	\end{center}
\end{figure}

The self energy $ \Sigma ^* (p)$ of fermions  satisfies the following Dyson equations  illustrated in Figure \ref{fig.5},
  \begin{eqnarray}
		i \Sigma ^* (p)&= & \frac{g^2}{(2\pi)^4}  \int d^4k \gamma _{\nu}S(p-k) \Gamma _{\mu}(p-k,p) D^{\mu\nu}(k)
		                                 \nonumber \\
		             &+&   \frac{1}{(2\pi)^4} \left( \frac{m_f}{\langle \widehat{ \beta}\rangle} \right )^2 \int d^4k S(p-k) \widehat{\Gamma}_H(p-k,p) G(k)   
		                                 \nonumber \\
		              &\equiv &    i\Sigma _1(p) + i\Sigma _2 (p)               
		 \label{eq:65}
\end{eqnarray}
where the vertex $\Gamma _{\mu}$ illustrated by a white triangle  is the vertex function between $B_{\mu}$ and $\psi$, and another vertex $\widehat{\Gamma} _H$ illusrated by a black triangle  is the vertex function between $H_{\beta}$ and $\psi$.  Correspondingly, the self energy $\Sigma (p)$ of fermion is composed of $\Sigma _1 (p)$ due to the coupling to $B_{\mu}$, and  $\Sigma _2 (p)$ due to the coupling to $H_{\beta}$ \cite{gold}.

Similarly,  the vacuum polarization $ \Pi_{\mu\nu}  ^*(p^2)$ of the massive gauge boson satisfies 
\begin{equation}
	i  \Pi_{\mu\nu}  ^*(p^2)=   \frac{g^2}{(2\pi)^4}  \int d^4k  tr[S(p+k) \gamma _{\mu}S(k) \Gamma _{\nu}(k,p+k)]  ,
	 	  						                	 \label{eq:66}
\end{equation}
and the vacuum polarization $ \Pi_{H}  ^*(p^2)$ of the Higgs-like  mode satisfies
\begin{equation}
	i  \Pi_{H}  ^*(p^2)=   \frac{g^2}{(2\pi)^4}  \int d^4k  tr[S(p+k) S(k) \widehat{\Gamma} _{H}(k,p+k)]    .
	 	  						                	 \label{eq:67}
\end{equation}

\subsection{The vertex  $\Gamma_{\mu} (p,p')$ between $B_{\mu}$ and $\psi$ }
\begin{figure}
	 	\begin{center}
\includegraphics [scale=0.4]{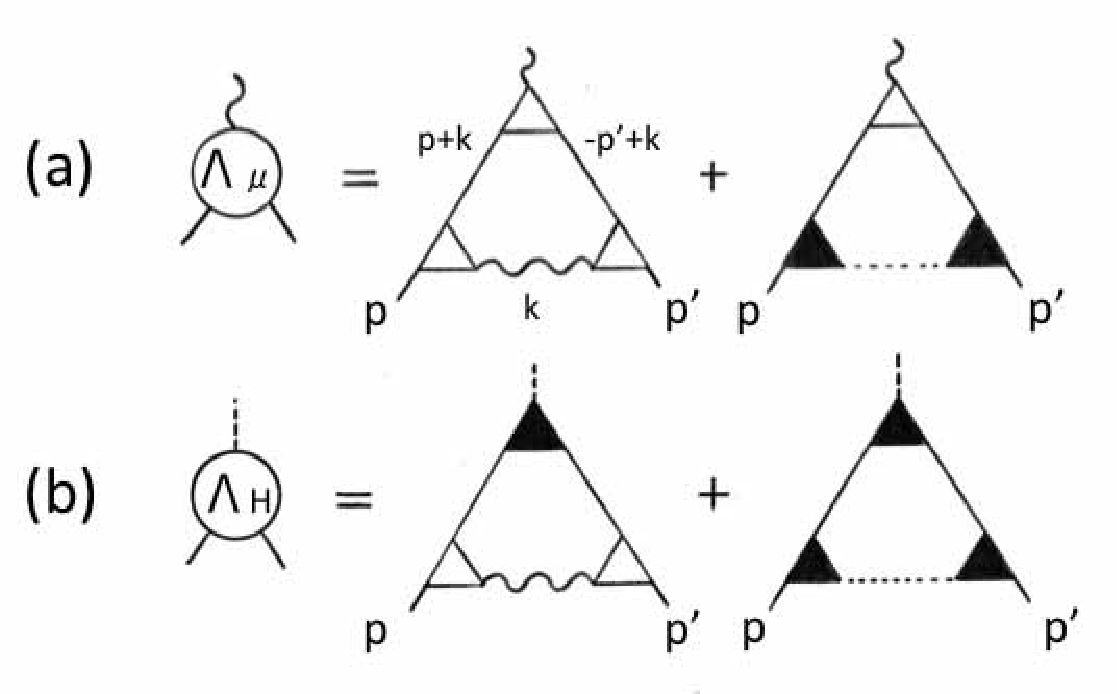}
\caption{(a) Proper vertex $\Lambda _{\mu}$ between the fermion $\psi$ and  the gauge field $B_{\mu}$, (b) Proper vertex $\Lambda _{H}$ between the fermion $\psi$ and  the Higgs-like  mode $H_{\beta}$.}
\label{fig.6}
    	\end{center}
\end{figure}

 For the vertex  $\Gamma_{\mu} (p,p')= \gamma_{\mu} +  \Lambda _{\mu}^* (p,p')$ between $B_{\mu}$ and $\psi$ in Eq.(\ref{eq:65}),   its proper vertex $ \Lambda _{\mu} ^* (p,p')$ satisfies the following Dyson equation as illustrated in Figure  \ref{fig.6}(a)
\begin{eqnarray}
	\lefteqn{	 \Lambda _{\mu}^* (p,p') }
	                     \nonumber      \\   
	&&=  \frac{g^2}{(2\pi)^4}  \int d^4k \Gamma _{\nu }(p+k, k) S(p+k) D^{\nu\rho}(k) 
	                               \nonumber \\
	&& \quad  \times \Gamma _{\rho}(k, k+p') S(-p'+k) 
		                   \Gamma _{\mu}(p+k, -p'+k) 
		                                 \nonumber \\
     &&+   \frac{1}{(2\pi)^4} \left( \frac{m_f}{\langle \widehat{ \beta}\rangle} \right )^2 \int d^4k \widehat{\Gamma }_{H }(p+k, k) S(p+k) G(k) 
		                                  \nonumber \\
      && \quad \times  \widehat{\Gamma} _{H} (k, k+p') S(-p'+k)
		                                \Gamma _{\mu}(p+k, -p'+k) 
		                                \nonumber \\
		             &&+ \cdots   .
		             		 \label{eq:68}
\end{eqnarray}
This $ \Lambda _{\mu}^* (p,p')$ is related to the fermion self-energy  $\Sigma _1^* (p)$ (due to the coupling to $B_{\mu}$) as \cite{war}
\begin{equation}
	  \frac{\partial}{\partial p_{\mu}}   \Sigma _1^* (p) =  \Lambda _{\mu} (p,p)    .
	 	  						                	 \label{eq:69}
\end{equation}

Similarly, if we consider a proper vertex $\Lambda _H(p,p')$  for   $\widehat{\Gamma} _H(p,p') = 1+\Lambda _H(p,p')$  between $H_{\beta}$ and $\psi $  in Eq.(\ref{eq:65}), it satisfies the  Dyson equation illustrated in Figure  \ref{fig.6}(b). The other fermion self-energy $\Sigma _2(p)$ (due to the coupling to $H_{\beta}$) is related to this $\Lambda _H$ as
\begin{equation}
	  \frac{\partial}{\partial p}   \Sigma _2 ^* (p) =  \Lambda_{H} (p,p)   .
	 	  						                	 \label{eq:825}
\end{equation}
With these $ \Lambda _{\mu} (p,p)$ and $ \Lambda _{H} (p,p)$,  we obtain the Green function of fermions 
\begin{equation}
	  [ S(p) ] ^{-1} = - \gamma \cdot p +m_f - \int_{p_0}^{p} dk^{\mu} \Lambda  _{\mu}(k,k)  -  \int_{p_0}^{p} dk  \Lambda  _{H}(k,k) .
	  						                	 \label{eq:70}
\end{equation}
We will introduce the following renormalization-constants $Z_i$ ($i=1\sim 5$ )
\begin{equation}
	  S(p) = Z_2 S^{ren} (p)   ,
	  						                	 \label{eq:71}
\end{equation}
\begin{equation}
	  D_{\mu\nu}(p^2) = Z_3 D_{\mu\nu}^{ren} (p^2)   ,
	  						                	 \label{eq:72}
\end{equation}
\begin{equation}
	  G(p^2) = Z_5 G^{ren} (p^2)   ,
	  						                	 \label{eq:725}
\end{equation}
\begin{equation}
	  \Gamma _{\mu}(p,p') = Z_1^{-1}  \Gamma_{\mu} ^{ren} (p,p')   ,
	  						                	 \label{eq:73}
\end{equation}
\begin{equation}
	 \widehat{ \Gamma} _{H}(p,p') = Z_4^{-1}  \widehat{\Gamma}_{H} ^{ren} (p,p')    .
	  						                	 \label{eq:735}
\end{equation}

Let us estimate the divergence appearing successively in the perturbation expansion  in Eq.(\ref{eq:68}).

(1) In the expansion of  $\Lambda _{\mu}(p,p')$,  we focus on $\Lambda _{\mu}$ of the order of $g^{2j}$ obtained   by $j$ times of iteration.   These  $\Lambda _{\mu}$ contain  $2j$ $S(p)$,  $j$ $D(p^2)$, and $2j$ $\Gamma _{\mu} (p,p')$, in addition to the original $\Gamma _{\mu}$. 
 When  $S(p)$, $D(p^2)$, $\Gamma_{\mu}$  and $\widehat{\Gamma} _H$  are replaced by their counterparts in Eqs.(\ref{eq:71}) $\sim$ (\ref{eq:735}),  this $\Lambda _{\mu} (p,p')$ is  renormalized as
\begin{equation}
	 \Lambda _{\mu} (p,p') =	(Z_1^{-1}  Z_2 Z_3^{1/2})^{2j} Z_1^{-1}  \Lambda _{\mu}^{ren}  (p,p')	  ,
	 			                	 \label{eq:736}
\end{equation}

(2) Next,  we focus on each $\Lambda _{\mu}$ of the order of  $(m_f/\langle \widehat{ \beta}\rangle)^{2j}$. In addition to the original $\Gamma _{\mu}$,  they contain $2j$ $S(p)$,  $j$ $G(p^2)$, and  $2j$ $\widehat{\Gamma} _{H} (p,p')$.  Hence,  this  $\Lambda _{\mu} (p,p')$ is  renormalized as
\begin{equation}
	 \Lambda _{\mu} (p,p') =	(Z_4^{-1}  Z_2 Z_5^{1/2})^{2j} Z_1^{-1}  \Lambda _{\mu}^{ren}  (p,p')	  ,
	 			                	 \label{eq:737}
\end{equation}
　If  $g$ is  replaced by
\begin{equation}
	 g^{ren} =	Z_1^{-1}  Z_2 Z_3^{1/2}  g	   ,		
	 			                	 \label{eq:74}
\end{equation}
and $\langle \widehat{ \beta}\rangle$ by
\begin{equation}
	 \langle \widehat{ \beta}\rangle _{ren} ^{-1}= Z_4^{-1}  Z_2 Z_5^{1/2}  \langle \widehat{ \beta}\rangle ^{-1} ,
	 			                	 \label{eq:7501}
\end{equation}
the total proper vertex is renormalized as
\begin{equation}
	 \Lambda _{\mu} (p,p') =	Z_1^{-1}  \Lambda _{\mu}^{ren}  (p,p', g^{ren},  \langle \widehat{ \beta}\rangle_{ren})	  ,
	 			                	 \label{eq:7601}
\end{equation}
For the higher-order terms with various combination of $g^2$ and $(m_f/\langle \widehat{ \beta}\rangle)^2$, similar renormalization is possible.

\subsection{The vertex  $\widehat{\Gamma} _H(p,p')$ between $H_{\beta}$ and $\psi $}
 For the vertex  $\widehat{\Gamma} _H(p,p') = 1+\Lambda _H(p,p')$ between $H_{\beta}$ and $\psi $  in Figure  \ref{fig.6}(b),   the proper vertex $\Lambda _H(p,p')$ has  a similar structure to  $\Lambda _{\mu}(p,p')$   if  $\Gamma _{\mu}$ for the external $B_{\mu}$ is  replaced by  $\widehat{\Gamma} _H$ for the external $H_{\beta}$.  Hence, using Eqs.(\ref{eq:74}) and (\ref{eq:7501}), 
 $\Lambda _{H} (p,p')$ is renormalized as
\begin{equation}
	 \Lambda _{H} (p,p') =	Z_4^{-1}  \Lambda _{H}^{ren}  (p,p', g^{ren},  \langle \widehat{ \beta}\rangle_{ren})	,
	 			                	 \label{eq:82}
\end{equation}
and $Z_1=Z_4$. Since  Eqs.(\ref{eq:69}) and (\ref{eq:825}) also  hold for the renormalized quantities, $Z_2=Z_1=Z_4$ is obtained.  Using Eqs.(\ref{eq:7601}) and  Eq.(\ref{eq:82}) in (\ref{eq:70}), we obtain  the total renormalized fermion self-energy $\Sigma^{ren} (p)=\Sigma^* _1 (p)+\Sigma^* _2 (p)$, and the renormalized $S^{ren}(p)$.

\subsection{The vacuum polarization $\Pi(p)$ of the massive gauge field $B_{\mu}$}
 For the vacuum polarization $\Pi^*(p)$ of $B_{\mu}$ in Eq.(\ref{eq:63}), the proper vertex $\Delta _{\mu} (p,p')$ satisfying 
\begin{equation}
	  \frac{\partial}{\partial p_{\mu}}   \Pi ^* (p) =  \Delta _{\mu} (p,p)    ,
	 	  						                	 \label{eq:77}
\end{equation}
is defined. This  $\Delta _{\mu} (p,p)$ is useful, because the Green's function $D (p^2) $ of  $B_{\mu}$ is expressed as
\begin{equation}
	  [ D (p^2) ] ^{-1} = -p^2 + m_B^2 - \int _{p'_0}^{p} dk^{\mu} \Delta _{\mu} (k, k)    .
	  	  						                	 \label{eq:78}
\end{equation}
 \begin{figure}
	 	\begin{center}
\includegraphics [scale=0.4]{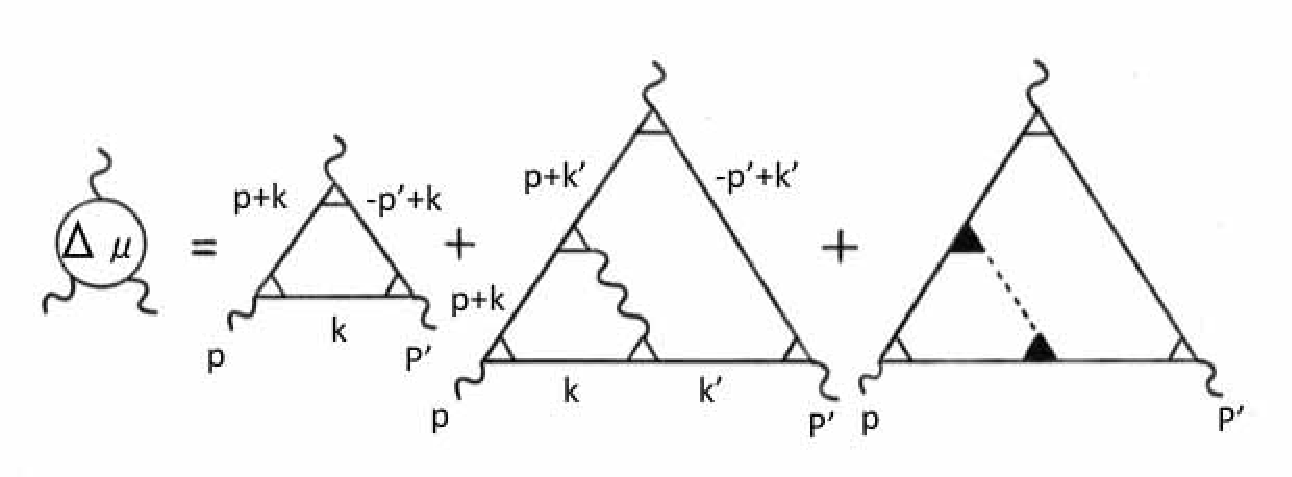}
\caption{Proper vertex $\Delta _{\mu}$ between   the gauge fields $B_{\mu}$.}
\label{fig.7}
    	\end{center}
\end{figure}
This  $ \Delta _{\mu} (p, p')$ satisfies the following Dyson equation illustrated in Figure \ref{fig.7} 
\begin{eqnarray}
		\lefteqn{ \Delta _{\mu} ^* (p,p')=  \frac{g^2}{(2\pi)^4}  \int d^4k 
		                                       \Gamma _{\nu }(p+k, k) S(p+k) S(k)}
		                                                  \nonumber \\
		             &&  \mbox{ } \times \Gamma _{\nu}(k, k+p') S(-p'+k) \Gamma _{\mu}(p+k, -p'+k) 
		                                 \nonumber \\
		             &&+   \frac{g^4}{(2\pi)^8}  \int d^4k d^4k' \Gamma _{\nu }(p+k, k) S(p+k) \Gamma _{\sigma}(p+k,p'+k)  
		                                   \nonumber \\
		               && \mbox{ } \times S(p+k') S(k) \Gamma _{\tau} (k, k') D^{\sigma\tau} (k- k') S(k')
		                                     \nonumber \\
		               && \mbox{ } \times \Gamma _{\nu}(k',-p'+k') S(k'-p') \Gamma _{\mu}(p+k',-p'+k')
		                                             \nonumber \\
		            &&+   \frac{1}{(2\pi)^8} g^2 \left( \frac{m_f}{\langle \widehat{ \beta}\rangle} \right )^2 \int d^4k d^4k' \Gamma _{\nu }(p+k, k) S(p+k) 
		                                            \nonumber \\
		            &&  \mbox{ } \times \widehat{\Gamma} _{H}(p+k,p'+k) S(p+k') 
		                                   \nonumber \\
		               && \mbox{ } \times S(k) \widehat{\Gamma} _{H} (k, k') G (k- k') S(k') \Gamma ^{\nu}(k',-p'+k') 
		                                    \nonumber \\
		               &&  \mbox{ } \times S(k'-p') \Gamma _{\mu}(p+k',-p'+k')
		                                             \nonumber \\      
		                 &&+  \cdots   .
		                               		             		 \label{eq:79}
\end{eqnarray}
Only few terms are written in the above expansion, but  it extends to the higher-order terms of $g^2$ and $(m_f/\langle \widehat{ \beta}\rangle)^2$. 

(1)  Let us consider $ \Delta _{\mu}$ of the order of  $g^{2j}$ in the above expansion.  They contain $(2j+1)$ $S(p)$,  $(j-1)$ $D_{\mu\nu}(p^2)$, and  $(2j+1)$ $\Gamma _{\mu} (p,p')$  \cite{re1}  
\begin{eqnarray}
	 \Delta _{\mu} (p,p')&=&  Z_2^{2j+1} Z_3^{j-1} Z_1^{-(2j+1)} \Delta^{ren} _{\mu} (p,p') 
	                                                     \nonumber \\
	                               &=& Z_2^{2j+1} Z_3^{j-1} Z_1^{-2j} Z_2^{-1} \Delta^{ren} _{\mu} (p,p')  
	                                                     \nonumber \\
	                               &=& (Z_1^{-1} Z_2 Z_3^{1/2} )^{2j} Z_3^{-1} \Delta^{ren} _{\mu} (p,p')      .
	                                                    \nonumber \\
	 	  						                	 \label{eq:795}
\end{eqnarray}
 Hence, if we replace $g$ by $g^{ren}$ defined in Eq.(\ref{eq:74}),  each $\Delta _{\mu}$ with $g^{2j}$ is renormalized.

(2)  Similarly, for $ \Delta _{\mu}$ with the coefficient $g^{i}(m_f/\langle \widehat{ \beta}\rangle)^j$,  it contains $(i+j+1)$ $S(p)$, $(i/2-1)D_{\mu\nu}(p)$, $j/2$ $G(p^2)$,   $(i+1)$ $\Gamma _{\mu} (p,p')$, and $j$ $\widehat{\Gamma} _{H} (p,p')$  \cite{re2}
 \begin{eqnarray}
	 \Delta _{\mu} (p,p') &=&  Z_2^{i+j+1} Z_3^{i/2-1} Z_5^{j/2} Z_1^{-(i+1)}  Z_4^{-j} \Delta^{ren} _{\mu} (p,p')  
	                                                    \nonumber \\
	 &=&    (Z_1^{-1} Z_2Z_3^{1/2} )^i (Z_4^{-1} Z_2 Z_5^{1/2} )^jZ_3^{-1} \Delta^{ren} _{\mu} (p,p') .
	                                           \nonumber \\
	 	  						                	 \label{eq:80}
\end{eqnarray}
 Hence, if we replace $g$ by $g^{ren}$ in  Eq.(\ref{eq:74}), and $\langle \widehat{ \beta}\rangle$ by $\langle \widehat{ \beta}\rangle_{ren}$ in Eq.(\ref{eq:7501}), each $\Delta _{\mu}$ with $g^i(m_f/\langle \widehat{ \beta}\rangle)^j$ is renormalized.  

(3) As a result, the total   proper vertex in Figure.\ref{fig.7} is renormalized as
\begin{equation}
	 \Delta _{\mu} (p,p') =	Z_3^{-1}  \Delta _{\mu}^{ren}  (p,p', g^{ren},  \langle \widehat{ \beta}\rangle_{ren})	.
	 			                	 \label{eq:81}
\end{equation}
Using Eq.(\ref{eq:81}) in Eq.(\ref{eq:78}),  we obtain the renormalized  $D^{ren}(p^2)$.

For the higher terms, similar renormalization is possible.
 \begin{figure}
	 	\begin{center}
\includegraphics [scale=0.4]{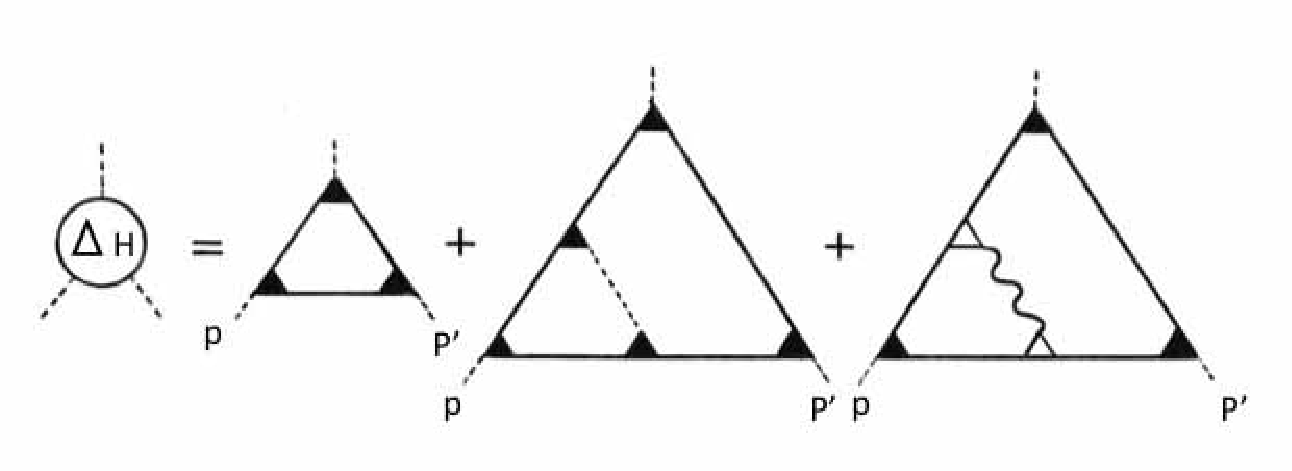}
\caption{Proper vertex $\Delta _{H}$ between the Higgs-like  modes $H_{\beta}$.}
\label{fig.8}
    	\end{center}
\end{figure}

\subsection{The vacuum polarization $\Pi _H(p)$ of the Higgs-like  mode $H_{\beta}$}
 For the vacuum polarization $\Pi _H(p)$ of the Higgs-like  mode $H_{\beta}$ in Eq.(\ref{eq:64}), the proper vertex $\Delta _{H} (p,p')$ satisfying
\begin{equation}
	  \frac{\partial}{\partial p}   \Pi ^* _H(p) =  \Delta _{H} (p,p)    ,
	 	  						                	 \label{eq:83}
\end{equation}
 is defined. This $\Delta _{H} (p,p)$ is useful, because the  Green's function $ G (p^2) $ of the Higgs-like  mode $H_{\beta}$ is expressed as
\begin{equation}
	  [ G (p^2) ] ^{-1} = -p^2 + m_H^2 - \int _{p''_0}^{p} dk \Delta _{H} (k, k)    .
	  	  						                	 \label{eq:835}
\end{equation}
The  proper vertex $\Delta_H(p,p')$ satisfies the Dyson equation shown in Figure \ref{fig.8}.  If  $\widehat{\Gamma} _H$ (black triangle) is replaced by $\Gamma _{\mu}$ (white triangle), and $\Gamma _{\mu}$ by  $\widehat{\Gamma} _H$,  it shows a similar structure to Figure \ref{fig.7} for $\Delta _{\mu}$. Hence,  in analogy with  Eq.(\ref{eq:81}),  the  proper vertex  is renormalized as
\begin{equation}
	 \Delta _{H} (p,p') =	Z_5^{-1}  \Delta _{H}^{ren}  (p,p', g^{ren},  \langle \widehat{ \beta}\rangle_{ren})	.
	 			                	 \label{eq:84}
\end{equation}
where $Z_5=Z_3$. Using Eq.(\ref{eq:84}) in Eq.(\ref{eq:835}),  we obtain the renormalized $G^{ren}(p^2)$.

\subsection{The renormalized masses of $\psi$, $B_{\mu}$ and $H_{\beta}$}
 The renormalized mass of the fermion $\psi$ is  a solution  of $[S(p)]^{-1}=0$,  in which $\Lambda ^{ren}_{\mu}(p,p', g^{ren}, \langle \widehat{ \beta}\rangle^{ren})$ and $\Lambda ^{ren}_{H}(p,p', g^{ren}, \langle \widehat{ \beta}\rangle^{ren})$  are used    in  Eq.(\ref{eq:70}), 
 
Similarly, the renormalized masses of  $B_{\mu}$ and $H_{\beta}$ are solutions  of $[D(p^2)]^{-1}=0$ in Eq.(\ref{eq:78}), and $[ G (p^2) ] ^{-1} =0$ in Eq.(\ref{eq:835}), respectively. (When such masses are calculated, $\Delta ^{ren}_{\mu}(p,p', g^{ren}, \langle \widehat{ \beta}\rangle^{ren})$ and $\Delta ^{ren}_{H}(p,p', g^{ren}, \langle \widehat{ \beta}\rangle^{ren})$ must be used, respectively.)

In view of  Eqs.(\ref{eq:62}) $\sim$  (\ref{eq:64}), we obtain the first approximation of such renormalized masses  as follows
\begin{equation}
	  m_f^{ren}= m_{f} - \Sigma ^{ren}(m_f)  ,
	  	  						                	 \label{eq:85}
\end{equation}
\begin{equation}
	  (m_B^{ren })^2= m^2_{B} - \Pi ^{ren}(m_B^2)   ,
	  	  						                	 \label{eq:86}
\end{equation}
\begin{equation}
	  (m_H^{ren })^2= m^2_{H} - \Pi _H^{ren}(m_H^2)   ,
	  	  						                	 \label{eq:87}
\end{equation}
in which $\Sigma ^{ren}$, $\Pi ^{ren}$ and $ \Pi _H^{ren}$ are self energy and vacuum polarizations in the right-hand side of  Eqs.(\ref{eq:70}),  (\ref{eq:78}) and  (\ref{eq:835}) using the renormalized quantities.

Since the Higgs Lagrangian density $L_h(x)$ does not exist in our model, all renormalization constants $Z_i$ ($i=$ 1 $\sim$ 5) are determined so as to absorb only the logarithmic divergence, not the quadratic one. 


\section{Discussion}

\subsection{Implications for the Higgs Lagrangian}
 The power of the Higgs Lagrangian density $L_h(x)$ in providing  experimental predictions  comes from its simple structure. Many quantities are derived from a single quantity $v_h$, such as the gauge boson's mass in $m_B= g v_h$, fermion's mass in $(m_f/v_h)\bar{\varphi} \varphi h$, and  the coupling of Higgs boson to gauge boson pair in $gv_h g^{\mu\nu}B_{\mu}B_{\nu} h$. If $L_h(x)$ is replaced by a more microscopic description, these $v_h$'s appearing in the different quantities may not  have the same value. The precise  measurement of the properties of the Higgs-like particle on this point will have a crucial  importance for future development.

 In the Higgs model, there are three  parameters: two coefficients $\mu$ and $\lambda$ in the Higgs potential $-\mu^2|h|^2+\lambda |h|^4$, and the fermion masses $m_f$. In the present model, there are five parameters:  four type of condensed  energies in the physical vacuum, (a) the two-point correlation in the physical vacuum leading to $m_B$  in  Eq.(\ref{eq:36}), (b) the condensed field-energy $\langle \widehat{ \beta}\rangle$ of  massless  gauge boson, leading to $m_f$  in  Eq.(\ref{eq:245}),  (c)  the three-point correlations in the physical vacuum leading to $V_h$ and $W_h$  in  Eqs.(\ref{eq:363}) and (\ref{eq:36314}). In addition to them, (d) the upper end of energy-momentum $\Lambda$ of fermions involved  in the excitation  in  Eq.(\ref{eq:46}).

 When the precise measurement of the Higgs-like particle is performed, the above degree of freedom will turn out to be important.

 \subsection{Extension to the electroweak interaction}
 When we apply the present model to  the electroweak interaction,  it is appropriate to begin with  the third generation, in which  fermions with  large masses, such as the  top and bottom quarks (plus $\tau$ lepton and $\tau$ neutrino), are included. Such a Lagrangian  density without the Higgs field is given by
\begin{eqnarray}
	L_0(x)	&=& -\frac{1}{4} B^{\mu\nu}B_{\mu\nu}  -\frac{1}{4} W^{a\mu\nu}W^a_{\mu\nu}
	                                 \nonumber \\   
	  &+&  \bar{q}(i\partial_{\mu} +g\tau_aW_{\mu}^a +g'Y_QB_{\mu}) \gamma ^{\mu}q
	                                    \nonumber \\
	   & +&  \bar{l}(i\partial_{\mu} +g\tau_aW_{\mu}^a +g'Y_LB_{\mu}) \gamma ^{\mu}l
	                                  \nonumber \\ 
	   &+&   \bar{r} \left( i\partial_{\mu} +g'
	                                 \left [ \begin{array}{cc} Y_t& 0 \\ 0&Y_b \end{array} \right ] B_{\mu} \right) \gamma ^{\mu}r
	                                   \nonumber \\
	    & +&  \bar{l}_r \left( i\partial_{\mu}  +g'
	                                       \left [ \begin{array}{cc} 0 & 0 \\ 0&Y_ {\tau} \end{array} \right ] B_{\mu} \right) \gamma ^{\mu}l_r     ,   
	                          		                						                	 \label{eq:622}
\end{eqnarray}
where 
\begin{eqnarray}
	 B^{\mu\nu} &=& \partial_{\mu}B^{\nu}-\partial_{\nu}B^{\mu}, 
	                             \nonumber \\
	   W_{\mu\nu}^{a} &=& \partial_{\mu}W^a_{\nu}-\partial_{\nu}W^a_{\mu} - gf^{abc}W^b_{\mu}W^c_{\nu}  ,
	                     	 	   	 \label{eq:633}
\end{eqnarray}
and massless fermions (top and bottom quarks,  $\tau$ lepton and $\tau$ neutrino) make up  left-handed doublets $q$ and $l$, and right-handed doublets $r$ and $l_r$,
\begin{equation}
q=  \left( \begin{array}{cc} \varphi _ t \\  \varphi _b \end{array} \right)_L, 
l=  \left( \begin{array}{cc}  \varphi  _{\nu_{\tau}}  \\  \varphi _ {\tau} \end{array} \right)_L,
r=  \left( \begin{array}{cc} \varphi _ t \\  \varphi _b \end{array} \right)_R,
l_r=  \left( \begin{array}{cc}  0  \\ \varphi _{\tau} \end{array} \right)_R
                                  \nonumber \\
                        \label{eq:634}
 \end{equation}
For the quarks in the third generation, the physical vacuum in  Eq.(\ref{eq:16})  is generalized so that it reflects $SU(2) \times U(1)$ symmetry of the top and bottom quarks  as
\begin{eqnarray}
\lefteqn{ | \widetilde{0} \rangle= e^{i\tau_3 \alpha _3(x)} }
                   \nonumber \\
 &&  \mbox{} \times 
        \left( \begin{array}{cc}   \prod_{p,s} \left( \cos \theta^{t}_{\mbox{\boldmath $p$}} 
		          + \sin \theta^{t}_{\mbox{\boldmath $p$}} e^{i\alpha_3(x)}b_{t}^{s\dagger}(-\mbox{\boldmath $p$}) a_{t}^{s\dagger}(\mbox{\boldmath $p$}) \right)  |0_r\rangle  \\
		             \prod_{p,s} \left( \cos \theta^{b}_{\mbox{\boldmath $p$}} 
		          + \sin \theta^{b}_{\mbox{\boldmath $p$}} e^{i\alpha_3(x)} b_{b}^{s\dagger}(-\mbox{\boldmath $p$}) a_{b}^{s\dagger}(\mbox{\boldmath $p$})   \right) |0_r\rangle
    \end{array} \right)
                                   \nonumber \\
                        \label{eq:645}
    \end{eqnarray}
In this case, the following condensed-energy accumulates in the physical vacuum.    The field energy  of the massless gauge boson  condenses in the vacuum as
\begin{eqnarray}
         &\widehat{g}^2&  \langle \widetilde{0}|  \int T^{00}_{W,(t)}(x)d^3x  |\widetilde{0}\rangle +\widehat{g'}^2 Y_Q \langle \widetilde{0}| \int T^{00}_{B,(t)} (x)d^3x  |\widetilde{0}\rangle
                                \nonumber \\
          = &\widehat{g' }^2&Y_t \langle \widetilde{0}| \int T^{00}_{B,(t)} (x)d^3x  |\widetilde{0}\rangle = m_t   ,
                                               	 	  						                	 \label{eq:646}
\end{eqnarray}
then producing the mass of the top quark.  Similarly,  the kinetic energy of the left-handed and right-handed quarks  separately condense in vacuum

\begin{equation}
	  m_W^2 =  g^2  \int  d^4Y \langle \widetilde{0} |  \bar{q}(Y) \tau _a \gamma_{\mu} q (Y)
                                                     \bar{q}(0)\tau _a  \gamma^{\mu} q (0)    |\widetilde{0} \rangle ,
                                                               \nonumber \\
	  	  						                	 \label{eq:66}
\end{equation}

and   
\begin{eqnarray}
		\lefteqn{ m_B^2 = g'^2  \int d^4Y } 
		            \nonumber \\
      && \mbox{}  \times \langle \widetilde{0} | 
	                          \left [\bar{q}(Y)Y_Q   \gamma_{\mu} q(Y) +  \bar{r}(Y) \left( \begin{array}{cc} Y_t& 0 \\ 0&Y_b \end{array} \right) 
	                                                           \gamma_{\mu} r (Y)\right ] 
	                                  \nonumber \\ 
	&&  \mbox{} \times  \left [\bar{q}(0)Y_Q   \gamma^{\mu} q(0) + \bar{r}(0) \left( \begin{array}{cc} Y_t& 0 \\ 0&Y_b \end{array} \right)  \gamma^{\mu}  r (0)\right ] 
	                                                           |\widetilde{0}   \rangle 	
	                                                            \nonumber \\
	  	                                                             		                	 \label{eq:67}
\end{eqnarray}
leading to masses of real gauge bosons.   
By this generalization,  the following possibilities are  expected.

(a) In the GWS model,   $W_{\mu} ^a$ and  $B_{\mu}$  are reorganized to $W_{\mu}^{\pm}, Z_{\mu}$ and $A_{\mu}$.  When $L_h(x)$ in  Eq.(\ref{eq:01}) is extended to include $W_{\mu}^a$ and $B_{\mu}$,  the couplings of $B_{\mu}$ to $W^1_{\mu}$ and  $W^2_{\mu}$ inevitably appear. In order to eliminate such a couplings, the vacuum condensate of Higgs particle is phenomenologically assumed to have a structure $\langle 0 | h | 0 \rangle =(0,v_h)$ . The extension of the present model  has a possibility of deriving the vanishing of $B_{\mu}W^1_{\mu}$ and $B_{\mu}W^2_{\mu}$, not from $(0,v_h)$,  but from the dynamical reason. 

(b) When the argument in Section 3 is extended across different generations, a microscopic explanation of CKM matrix is expected. If $a(p)$ and $b(-p)$ in the second term of  Eq.(\ref{eq:17})  represent the down and strange quarks respectively, an orthogonal transformation  giving rise to the mixing of different generations is introduced for the state coupled to the gauge field, and  the Cabibbo angle is defined  as an angle of such a transformation.

(c) In the electroweak version of the present model,   two types of condensed kinetic energy of massless quarks in   Eqs.(\ref{eq:66}) and  (\ref{eq:67}) are assumed, and therefore the definition of Weinberg angle $\theta _W$, which determines the mixing of $W^3_{\mu}$ and $B_{\mu}$, will be slightly changed.

(d)   Intensely examined by experiments now are the custodial symmetry in  the coupling of the observed  Higgs-like particle to $A_{\mu}$, $Z_{\mu}$ and $W_{\mu}^a$, and the existence  of the anomalous  coupling in it \cite{ano}.   In the Higgs model, the coupling of the Higgs field $h$  to $A_{\mu}$, $Z_{\mu}$ and $W_{\mu}^a$ comes from the common $v_h$.  In the electroweak version of the present model, the effective coupling of the Higgs-like  mode to $A_{\mu}$, $Z_{\mu}$ and $W_{\mu}^a$  have a variety of strength and $q^2$-dependence.  The  anomalous effective coupling such as  Eq.(\ref{eq:925}) is  worth precise measurements.

(e)  When the present model is generalized to the electroweak interaction,  the   Higgs model  and the present model will give different predictions in the intermediate- and high-energy processes. For example,  the present model predicts the  total cross section $\sigma(q^2)$ of the  fermion-antifermion annihilation to gauge boson pair,  in a different way from  the Higgs model.  In addition to the rise at $q^2=(2m_B)^2$, the gradual increase and decrease of $\sigma(q^2)$  around $q^2=(2m_f)^2$ is expected.
 The precise measurements of  properties of the recently discovered Higgs-like particle are expected  in future experiment.  \newline 

\appendix
\section{Contributions of the space-like path to amplitudes}
The motion  in  Figure.\ref{fig.1}(a) is  expressed by the second term in the following  amplitude
 \begin{eqnarray}
		\lefteqn{ Amp (t_2-t_1)= 1 -  \int \frac{ d\mbox{\boldmath $p$}'}{(2\pi)^3 \omega_{p'}} 
                                                                  \int d\mbox{\boldmath $x$}_1d\mbox{\boldmath $x$}_2}
		                                \nonumber \\
		   && \times   \langle 0| b^{\dagger }(\mbox{\boldmath $x$}_2) U_2(\mbox{\boldmath $x$}_2) 
		            U_1(\mbox{\boldmath $x$}_1) a(\mbox{\boldmath $x$}_1) |0\rangle
		                          \nonumber \\
		   && \times  \exp ( i [ \mbox{\boldmath $p$}'\cdot (\mbox{\boldmath $x$}_2-\mbox{\boldmath $x$}_1)-\omega _{p'}(t_2-t_1)  ] ), 
				                	 \label{eq:511}
\end{eqnarray} 
where $\omega _{p'}= | \mbox{\boldmath $p$}'|$ for massless fermions.  
Rewriting the differential $d\mbox{\boldmath $p$}'$ in Eq.(\ref{eq:511}) by $\omega _{p'}^2d\Omega d\omega _{p'}$,
this is an example of such a type of  Fourier integral
\begin{equation}
f(t)=\int^{\infty}_{0}	F(\omega) \exp(i\omega t) d\omega .
				                	 \label{eq:512}
\end{equation} 
Here, for an arbitray $F(\omega)$,  $f(t)$ cannot be zero for any finite interval of $t$.  Hence, even if $\mbox{\boldmath $x$}_2$  is outside the light cone of  $\mbox{\boldmath $x$}_1$   in Eq.(\ref{eq:511}),  the integral  is not zero. The amplitude  in Eq.(\ref{eq:511}) is determined  not only by the timelike, but also by  the spacelike paths  \cite{fey}.  

\section{ The proof of  $A^s(\mbox{\boldmath $p$})|\widetilde{0}\rangle =B^s(-\mbox{\boldmath $p$})|\widetilde{0}\rangle =0$}
 The proof of  $A^s(\mbox{\boldmath $p$})|\widetilde{0}\rangle =B^s(-\mbox{\boldmath $p$})|\widetilde{0}\rangle =0$   for  Eq.(\ref{eq:135}) is  as follows. Defining an operator
\begin{equation}
                 K= i \sum_{p,s}  \theta_{\mbox{\boldmath $p$}} [b^{s\dagger}(-\mbox{\boldmath $p$})a^{s\dagger}(\mbox{\boldmath $p$})-a^s(\mbox{\boldmath $p$})b^s(-\mbox{\boldmath $p$})] ,
						                	 \label{eq:951}
\end{equation}
and applying the following expansion  
\begin{equation}
                    e^{-iK} F e^{iK} = F + [-iK,F]+  \frac{1}{2!} \left[-iK, [-iK,F] \right] + \cdots ,
						                	 \label{eq:10}
\end{equation}
 to the operators $a^s(\mbox{\boldmath $p$})$ and $b^s(-\mbox{\boldmath $p$})$ for $F$,  and to  Eq.(\ref{eq:951}) for $K$,  we  rewrite Eqs.(\ref{eq:8}) and (\ref{eq:9}) in the following compact form
\begin{equation}
		A^s(\mbox{\boldmath $p$})= e^{-iK}a^s(\mbox{\boldmath $p$})e^{iK}    , \quad					            
		B^s(-\mbox{\boldmath $p$})= e^{-iK}b^s(-\mbox{\boldmath $p$})e^{iK}   .
						                	 \label{eq:12}
\end{equation}
The vacuum  $|\widetilde{0}\rangle$ in  $A^s(\mbox{\boldmath $p$})|\widetilde{0}\rangle =B^s(-\mbox{\boldmath $p$})|\widetilde{0}\rangle =0$  is expressed as 
\begin{eqnarray}
		|\widetilde{0}\rangle&=& e^{-iG}|0\rangle
		                                    \nonumber \\
		 &=&\exp\left( \sum_{p,s}\theta_{\mbox{\boldmath $p$}} [b^{s\dagger}(-\mbox{\boldmath $p$})a^{s\dagger}(\mbox{\boldmath $p$})
		                                        -a^s(\mbox{\boldmath $p$})b^s(-\mbox{\boldmath $p$}) ]\right ) |0\rangle  .
		                                         \nonumber \\
         	&=& \prod_{p,s} \left[ \sum_{n} \frac{1}{n!}  \theta_{\mbox{\boldmath $p$}}^n   [b^{s\dagger}(-\mbox{\boldmath $p$})a^{s\dagger}(\mbox{\boldmath $p$})
		                                        -a^s(\mbox{\boldmath $p$})b^s(-\mbox{\boldmath $p$})]^n \right]    |0\rangle.    
		                                                       \nonumber \\
		                                           		                                                 \label{eq:13}
\end{eqnarray}
Because  massless fermions and antifermions  obey Fermi statistics,  only a single particle can occupy each state on the hyperboloid  ($p^2=0$)  set at each point in space, and we obtain  for each $\mbox{\boldmath $p$}$
\begin{eqnarray}
	\lefteqn{\sum_{n} \frac{\theta^n}{n!} (b^{\dagger}a^{\dagger}-ab)^n |0\rangle } 
	                                                    \nonumber \\
	&&= |0\rangle + \theta b^{\dagger} a^{\dagger}  |0\rangle - \frac{\theta^2}{2!} abb^{\dagger} a^{\dagger}  |0\rangle 
		                             -  \frac{\theta^3}{3!} b^{\dagger} a^{\dagger} abb^{\dagger} a^{\dagger} |0\rangle 
		                                          \nonumber \\
		 &&+ \frac{\theta^4}{4!} abb^{\dagger} a^{\dagger} abb^{\dagger} a^{\dagger} |0\rangle  + \cdots  .
						                	 \label{eq:1333}
\end{eqnarray}
In this expansion,  $\cos \theta_{\mbox{\boldmath $p$}}$ appears in the sum of even-order  terms of $\theta$,  and $\sin \theta_{\mbox{\boldmath $p$}}$ appears  in  the sum of  odd-order terms, and then Eq.(\ref{eq:135}) is yielded.

\section{The  anomalous coupling of Higgs-like   mode }
  In the anomalous effective coupling of the Higgs-like collective mode to gauge boson,   $F_2(q^2,p^2,k^2, m_f)$ and $F_3(q^2,p^2,k^2, m_f)$ are included  in  Eq.(\ref{eq:925}). Following the standard procedure, we obtain such $F_2$ and $F_3$ as
\begin{eqnarray}
	 \lefteqn{F_2(q^2,p^2,k^2, m_f) 
	    = - \frac{16 m_B^2}{(4\pi)^2} 
	                \left[   \frac{ q^4-3(p^2+k^2)q^2 + 8p^2k^2}{ q^2(q^2-4p^2) (q^2-4k^2) } \right] }
	                                     \nonumber \\
	  &&   - \frac{16 m_B^2}{(4\pi)^2}  \frac{(p^2+k^2)q^6 + 2(p^4+k^4-8p^2k^2)q^4 + 64p^4k^4}{ q^2(q^2-4p^2)^2 (q^2-4k^2)^2 } 
	                                      \nonumber \\
	  &&\quad  \mbox{} \times f \left(\frac{q}{2m_f} \right)   
	                                           \nonumber \\
	  && +  \frac{16m_B^2}{(4\pi)^2} \left[  \frac{(q^2+2p^2)p^2}{q^2(q^2-4p^2)^2}  f  \left(\frac{p}{2m_f} \right)  + (p \leftrightarrow k) \right]
	                                                     \nonumber \\
	&& +  \frac{4}{(4\pi)^2}    \frac{ (q^2-2p^2) (q^4 -[6p^2+4m_f^2]q^2 -4p^4+16p^2m_f^2 ) }{q^2(q^2-4p^2)^2}
	                                           \nonumber \\
	&& \quad \mbox{}  \times  C_0(p^2, p^2, q^2, m_f^2, m_f^2, m_f^2)  
	                                              \nonumber \\
	 && + (p \leftrightarrow k) ,
	                            		                	 \label{eq:566}
\end{eqnarray}
\begin{eqnarray}
	 \lefteqn{F_3(q^2,p^2,k^2, m_f) 
	    =  \frac{16 m_B^2}{(4\pi)^2} 
	                \left[   \frac{ (p^2+k^2)q^2 - 8p^2k^2}{ q^2(q^2-4p^2) (q^2-4k^2) } \right] }
	                                     \nonumber \\
	  && + \frac{32 m_B^2}{(4\pi)^2} \left[  \frac{(p^2+k^2)q^6 - (p^4+k^4+ 16p^2k^2)q^4 }{ q^2(q^2-4p^2)^2 (q^2-4k^2)^2 } \right]
	                              f \left(\frac{q}{2m_f} \right)   
	                                           \nonumber \\
	   && + \frac{32 m_B^2}{(4\pi)^2} \left[  \frac{( 24(p^4k^2+p^2k^4)q^2 - 32p^4k^4}{ q^2(q^2-4p^2)^2 (q^2-4k^2)^2 } \right]
	                              f \left(\frac{q}{2m_f} \right)   
	                                           \nonumber \\                                         
	  && -  \frac{32 m_B^2}{(4\pi)^2}  \left[ \frac{(q^2 - p^2)p^2}{q^2(q^2-4p^2)^2}  f  \left(\frac{p}{2m_f} \right)  + (p \leftrightarrow k)\right]
	                                                     \nonumber \\
	&& +  \frac{8}{(4\pi)^2}    p^2 \frac{ (q^4 + [4m_f^2 - 2p^2]q^2 + 4p^4 - 16p^2m_f^2 ) }{q^2(q^2-4p^2)^2}
	                                              \nonumber \\
	 && \quad  \times C_0(p^2, p^2, q^2, m_f^2, m_f^2, m_f^2)  
	                                               \nonumber \\
	  && + (p \leftrightarrow k)   .
	                             						                	 \label{eq:567}
\end{eqnarray}


\begin{thebibliography}{99}
\bibitem{coh}
An example of this viewpoint  is the coherent state of photons. Photons coupled to the macroscopic classical current form the  coherent state. The macroscopic classical current produces the sequential  perturbation to photons, and the coherent state is the eternal intermediate state.
\bibitem{nuc}
In nuclear physics,  the  relativistic many-body states of massive fermions have been studied. However, they are  many-body states after spontaneous symmetry breaking.
\bibitem{law}
This is necessary even under the law with time-reversal-symmetry.
\bibitem{non} 
In the non-relativistic physics,  because the reversal of temporal order  does not occur, the  state given by $a^s(\mbox{\boldmath $p$}) |0\rangle=b^{s\dagger }(-\mbox{\boldmath $p$})|0\rangle =0$ is always the free vacuum.
\bibitem{cry} 
The  massive particles in the symmetric vacuum  have their own phase freely. In contrast, the massless fermions and antifermions  in $|\widetilde{0}\rangle$ have a common phase $\alpha(x)$, and can not have their own phase freely.  This is a kind of symmetry breaking of the system,  in the same sense that the translational  symmetry in the gas or liquid  is broken in the crystal lattice.
\bibitem{sta}
 $\cos ^2\theta_{\mbox{\boldmath $p$}}$ and $\sin ^2\theta_{\mbox{\boldmath $p$}}$  in Eq.(\ref{eq:23}) were  derived  in  \cite{nam} as inner products of the Lorentz-boosted 2-component  massive spinor $u$, $v$  with that of the massless ones. The  derivation   from  the stability condition of the physical vacuum  implies that the massive fermion and antifermion  are not only    Lorentz-covariant but also  stable constructions of  the massless ones.
\bibitem{qed}
Although the fermions  in the bubble diagram in Figure \ref{fig.15}  and the triangle loop in  \ref{fig.85}(a) are  massless, its excitation from $|\widetilde{0}\rangle$ is simply written using the massive fermion operator  $\psi(x)$.  Hence, $m_f$ appears in the denominators in Eqs.(\ref{eq:46}) and (\ref{eq:51}).
\bibitem{gold}
 The Goldstone mode $\phi(x)$  also couples to fermion $\psi(x)$ in $L_3(x)$ of Eq.(\ref{eq:50}). But, when  the  propagator $D^{\mu\nu}(q)$  in Eq.(\ref{eq:384}) is decomposed as 
 \begin{equation}
       D^{\mu\nu}(q) = \frac{-i}{q^2-m_B^2} \left( g^{\mu\nu} - \frac{q^{\mu}q^{\nu}}{m_B^2} \right) +  \frac{-i}{q^2}\frac{q^{\mu}q^{\nu}}{q^2} , \nonumber \\
 \end{equation}
  the renormalization of the fermion self-energy $\Sigma(p)$ due to the propagator of $\phi(x)$  is cancelled by that due to the second term  of this $D^{\mu\nu}(q)$. Hence, only the renormalization of  
  $\Sigma(p)$ due to $D^{\mu\nu}(q)$ is considered.
\bibitem{re1}
Since one of $(2j+1)$ $\Gamma _{\mu} (p,p')$ comes from the differential of $S^{-1}(p)$,  $Z^{-(2j+1)}_1$ in the first line of Eq.(\ref{eq:795}) must be replaced by  $Z^{-2j}_1Z_2^{-1}$.  
 For  the perturbation expansion of $\Pi_H$ in Figure \ref{fig.8} as well,  the same  situation occurs  for  $(2j+1) \widehat{\Gamma} _{H} (p,p')$.
\bibitem{re2}
 For  the order of $g^2(m_f/\langle \widehat{ \beta}\rangle)^2$ in  Eq.(\ref{eq:79}) where $i=2$ in $g^i(m_f/\langle \widehat{ \beta}\rangle)^j$,  $D_{\mu\nu} $ is not explicitly written. But,  for the higher-order terms in the expansion,  $(i/2-1)D_{\mu\nu}$ must be added.  Hence, $Z_3^{i/2-1}$ appears in the first line of Eq.(\ref{eq:80}).  As in \cite{re1}, since one of $(i+1)$ $\Gamma _{\mu} (p,p')$ comes from the differential of $S^{-1}(p)$,  $Z^{-(i+1)}_1$ must be replaced by  $Z^{-i}_1Z_2^{-1}$ in the second line.
\end{thebibliography}
\end{document}